\documentclass[%
 pra, twocolumn,
 superscriptaddress,
 showpacs,
 longbibliography,
 aps,
 showkeys]{revtex4-2}

\usepackage[pdftex]{hyperref,color,graphicx}
\usepackage{amsfonts,amssymb,amsmath}
\usepackage[separate-uncertainty=false]{siunitx}
\usepackage[english]{babel}
\usepackage[utf8]{inputenc}
\usepackage[T1]{fontenc}
\usepackage[toc,page]{appendix}
\usepackage{graphicx}

\usepackage[space]{grffile}
\usepackage{latexsym}
\usepackage{textcomp}
\usepackage{longtable}
\usepackage{multirow,booktabs}
\usepackage{url}
\hypersetup{colorlinks=false,pdfborder={0 0 0}}
\usepackage[english]{babel}
\usepackage{dcolumn}
\usepackage{bm}
\usepackage{units}
\usepackage{upgreek}
\usepackage{color}
\usepackage{cancel}
\usepackage{hyperref}
\usepackage{braket}
\usepackage{epstopdf}

\newcommand{\bbN}{\mathbb{N}}

\begin{document}

\title{Light-shift induced behaviors observed in momentum-space quantum walks}

\author{Nikolai Bolik}
\affiliation{Institute of Theoretical Physics, Heidelberg University, Philosophenweg 16, 69120 Heidelberg, Germany}

\author{Caspar Groiseau}
\affiliation{Departamento de F\'isica Teórica de la Materia Condensada and Condensed Matter Physics Center (IFIMAC), Universidad Autónoma de Madrid, 28049 Madrid, Spain}

\author{Jerry H. Clark}
\affiliation{Department of Physics, Oklahoma State University, Stillwater, Oklahoma 74078-3072, USA}

\author{Alexander Gresch}
\affiliation{Quantum Technology Research Group, Heinrich Heine University D\"usseldorf, Universit\"atsstr. 1, 40225 D\"usseldorf, Germany}

\author{Siamak Dadras}
\affiliation{TOPTICA Photonics Inc., 5847 County Road 41, Farmington, New York 14424, USA}

\author{Gil S. Summy}
\affiliation{Department of Physics, Oklahoma State University, Stillwater, Oklahoma 74078-3072, USA}
\affiliation{Airy3D, 5445 Avenue de Gasp\'e Suite 230, Montr\'eal, Qu\'ebec H2T 3B2, Canada}

\author{Yingmei Liu}
\email{yingmei.liu@okstate.edu}
\affiliation{Department of Physics, Oklahoma State University, Stillwater, Oklahoma 74078-3072, USA}

\author{Sandro Wimberger}
\email{sandromarcel.wimberger@unipr.it}
\affiliation{Dipartimento di Scienze Matematiche, Fisiche e Informatiche, Universit\`{a} di Parma, Parco Area delle Scienze 7/A, 43124 Parma, Italy}
\affiliation{INFN, Sezione di Milano Bicocca, Gruppo Collegato di Parma, Parco Area delle Scienze 7/A, 43124 Parma, Italy}

\begin{abstract}
Over the last decade there have been many advances in studies of quantum walks (QWs) including a momentum-space QW recently realized in our spinor Bose-Einstein condensate system. This QW possessed behaviors that generally agreed with theoretical predictions, however, it also showed momentum distributions that were not adequately explained by the theory.  We present a novel theoretical model, which proves that the coherent dynamics of the spinor condensate is sufficient to explain the experimental data without invoking the presence of a thermal cloud of atoms as in the original theory. Our numerical findings are supported by an analytical prediction for the momentum distributions in the limit of zero-temperature condensates. This current model provides more complete explanations to the momentum-space QWs that can be applied to study quantum search algorithms and topological phases in Floquet-driven systems.
\end{abstract}


\maketitle

\section{Introduction}
\label{sec-intro}

Quantum walks (QWs) have been under intensive investigation over the last two decades since they can outrun
classical algorithms for many practical problems \cite{Kempe2003, Portugal, Kendon}. For example, the Grover search algorithm may be viewed as a quantum walk algorithm \cite{Portugal}. Due to quantum interference of various passes during  quantum walks, they exhibit quite different features when compared to their classical counterpart for which, in contrast, randomness and stochasticity play a crucial role \cite{Kempe2003}.
Similar to classical random walks there are essentially two types of quantum analogues, discrete-time and continuous-time quantum walks. In contrast to the latter, an additional coin-degree-of-freedom characterizes the former, where the state of the coin determines the walker's direction in the next step.

We apply a novel theoretical model to the discrete-time quantum walk implemented in our previous works~\cite{Dadras2018,Clark2021,dadras2019experimental} with spinor Bose-Einstein condensates (BECs), consisting of $^{87}$Rb atoms with an internal spin-1/2 degree of freedom.
In contrast to most other experimental realizations ~\cite{preiss2015strongly,dur2002quantum, eckert2005one, steffen2012digital, groh2016robustness,karski2009quantum,
chandrashekar2006implementing, travaglione2002implementing,
zahringer2010realization, schmitz2009quantum, perets2008realization,
peruzzo2010quantum, cardano2017detection, chen2018observation, tang2018experimental, poulios2014quantum, schreiber2010photons}, this QW occurs in quantized momentum space due to time-periodic kicks applied to the condensate. The experiments in Refs.~\cite{Clark2021,Dadras2018,dadras2019experimental} used the two ground-state Zeeman sublevels $ \ket{F = 1, m_{F} = 0} $ and $ \ket{F = 2, m_{F} = 0} $ of a Rubidium BEC to form an effective spin-$\frac{1}{2}$ system. 
The BEC is periodically subjected to pulses of standing-wave light generated by a laser tuned between the two ground states and a third excited level. The underlying description is that of the atom-optics kicked rotor (AOKR) as described in Refs.~\cite{Raizen1999, SW2011}, whose Hamiltonian is
\begin{equation}
    \expandafter\hat {\mathcal{H}} = \frac{1}{2}\expandafter\hat p^2 + k\mathrm{cos}(\expandafter\hat \theta)\sum\limits_{j=-\infty}^{\infty}\delta(t-j\tau).
    \label{eq1}
\end{equation}
Here, $\hat p$ and $\hat \theta$ represent the momentum and (angular) position operators, respectively, while $k$ is the strength of the laser kick and $\tau$ the time delay between consecutive pulses. Since the experiment is performed in a periodic lattice potential, we resort to Bloch's theorem to arrive at the angle description above. This necessitates the introduction of a dimensionless quasi-momentum $\beta \in [0,1)$. The width of the Gaussian-like quasi-momentum distribution is experimentally given by the initial temperature of the BEC, where e.g. a BEC at zero temperature would correspond to a fully resonant system with $\beta = 0$ for all atoms.
The typical value of the width of the $\beta$-distribution in our experimental system is of the order of a few percent in the Brillouin zone, i.e., $\beta_{\rm FWHM} \approx 0.025$.

The evolution during one period $\tau$ is then described by the following Floquet operator 
\begin{equation}
    \hat{\mathcal{U}} = \hat{\mathcal{U}}_{\rm f}\hat{\mathcal{U}}_{\rm k} = e^{-i\tau \frac{\hat{p}^2}{2}} e^{-i\hat \sigma_z k\cos(\hat{\theta})},
\label{eq2}
\end{equation}
which factorizes into a free evolution $\hat{\mathcal{U}}_{\rm f}$ and kick operator $\hat{\mathcal{U}}_{\rm k}$. Since $p = n + \beta$, with integer (quantized angula) momenta $n$, the free evolution equals the identity in quantum resonance conditions, i.e., for an evolution corresponding to a full Talbot time $\tau=4\pi$ and $\beta=0$. Under these resonance conditions,
the atoms move ballistically in momentum space, i.e., their momenta increase linearly with the number of applied kicks~\cite{WGF2003, SW2011}.

Because the kicking laser is detuned exactly between the two internal ground states \cite{Dadras2018, dadras2019experimental}, the potential felt by the two states is identical in size but opposite in sign, which reflects the $\hat \sigma_z$ Pauli matrix. The latter fact models a quantum walk whose direction in each step depends on the internal coin state. There is an important difference between our AOKR quantum walk and an ideal quantum walk as defined, e.g., in Ref.~\cite{Kempe2003}. In the latter at any step of the walk a certain position of the walker only couples to the nearest-neighbor positions, whilst in the AOKR quantum walks the coupling to other momentum classes is given by matrix elements which are Bessel functions of first kind~\cite{dadras2019experimental}. A priori, both internal states would see the same evolution due to the kicks, i.e., they would move symmetrically under the AOKR evolution.
To break this symmetry in the coupling, we use a ratchet effect imposed by an appropriate choice of the initial condition in the walker's space. Those ratchet states are a superposition of at least two neighboring momenta with a relative phase of $e^{i\pi/2}$, i.e.,
\begin{equation}
    \ket{\psi_{\rm R}} = \frac{1}{\sqrt{S}}\sum_s e^{is\pi/2} \ket{n=s},
    \label{eqratchet}
\end{equation}
where $S$ is the total number of involved momentum classes denoted by $s$. Such initial states can be generated experimentally via Bragg pulses~\cite{Ni2016, Ni2017}. The mean momentum transfer to individual states depends on the sign of the kicking potential that is different for the two internal states, as shown by $\hat{\mathcal{U}}_{\rm k}$ in Eq.~\eqref{eq2} \cite{Mark2007, Gil2008, SW2011}. It turns out to be of crucial importance that the larger the number $S$ in 
Eq.~\eqref{eqratchet} the less dispersion in the directed kicking occurs \cite{Ni2016, Ni2017}. Hence, the best correspondence to an ideal quantum walk is found for large $S \geq 3$, whilst for $S=2$ differences from ideal walks are visible in the central part of the walker's probability distribution \cite{SW2016}.

The coin operator is realized by a Rabi coupling between the two internal states of the atoms. This coupling is mediated by resonant microwave (MW) pulses, inducing a
unitary rotation on the Bloch sphere given by
\begin{equation}
\hat M(\alpha, \chi) = 
 \begin{pmatrix}
\cos(\frac{\alpha}{2})            & e^{-i \chi}\sin(\frac{\alpha}{2})  \\ 
-e^{i \chi}\sin(\frac{\alpha}{2}) & \cos(\frac{\alpha}{2}) \end{pmatrix},
 \label{eq-matrix} 
\end{equation}
where $\alpha$ and $\chi$ are real angles. 
An additional $\hat \sigma_z$ rotation is implementable by an accessible third angle that was not considered in Refs.~\cite{SW2016, Dadras2018, dadras2019experimental} and will also not be considered in this paper.
The experimental QWs in Ref.~\cite{Dadras2018,dadras2019experimental,Clark2021} were described by the following sequence of unitary operations 
\begin{equation}
    \hat{\mathcal{U}}_\mathrm{step}^j = [\hat{\mathcal{U}}\hat{Y}]^j \hat{\mathcal{U}}\hat{W},
\label{eq-walk}
\end{equation}
realizing $j\in \bbN$ steps of the walk applied to an initial state expressed by Eq.~\eqref{eqratchet}. Here 
\begin{equation}
    \hat{W} = \hat{M}\left(\frac{\pi}{2},0\right) = \frac{1}{\sqrt{2}}\begin{pmatrix}
1 & 1  \\ 
-1 & 1\end{pmatrix}
\label{eq-coin1}
,\end{equation}
and 
\begin{equation}
    \hat{Y} = \hat{M}\left(\frac{\pi}{2},-\frac{\pi}{2}\right) = \frac{1}{\sqrt{2}}\begin{pmatrix}
1 & i  \\ 
i & 1\end{pmatrix}
\label{eq-coin2}
\end{equation}
are two different coins that initialize and execute the walk, respectively. It is important that the two coins must be different in order to guarantee a symmetric evolution of the walker (see Ref.~\cite{Kempe2003}). 
The kick strength on the order of $ k\approx1.5 $ proves to resemble well an ideal walk with only nearest neighbor couplings \cite{SW2016,dadras2019experimental,Dadras2018,Clark2021}.
%The kick strength was chosen of the order $k\approx 1.5$ such that the evolution resembles best an ideal walk with only nearest-neighbor coupling \cite{SW2016,dadras2019experimental}.
For example, the experiments reported in Refs.~\cite{Dadras2018, dadras2019experimental, Clark2021} used $k=1.2$, $k=1.45$, and $k=1.8$. After $j$ steps, the momentum distribution of both internal states 
is measured using the standard absorption imaging procedure to yield the final observable $P(n,j)=P_{|1\rangle}(n,j)+P_{|2\rangle}(n,j)$. Note that all the experimental realizations so far implemented walks with only $S=2$, e.g., an initial ratchet state of the form
\begin{equation}
 \ket{\psi_{\rm R}} = \frac{1}{\sqrt{2}} \left( \ket{n=0}+i\ket{n=1} \right).
 \label{eq-rat-2}
\end{equation}
 
Numerical simulations of the walk given by Eq.~\eqref{eq-walk} showed a good resemblance to the ideal quantum walk~\cite{SW2016}, with ballistically moving sidepeaks and little probability at the center around $n=0$. However, the experiments observed a large non-vanishing part of the momentum distribution that stayed close to $n=0$ throughout the entire evolution of up to $j=15$ steps~\cite{Dadras2018, dadras2019experimental, Clark2021}. This observation was initially explained in Ref.~\cite{Dadras2018} by a rather large residual thermal atomic cloud that would make up about $ 10\% $ to $ 15\% $ of all the measured atoms. A thermal cloud would correspond to much hotter atoms uniformly distributed across the entire Brillouin zone ${\beta \in [0,1)}$. 
All non-resonant quasi-momenta ($\beta \neq 0$) essentially do not respond to the kicks and hence will move little and not at all contribute to the expected ballistic flanks in the distribution. In this paper we suggest a more complete theoretical interpretation of the experimental data, not involving a thermal cloud but based on the concurrence of a sequence of effects that resulted in a deviation of the experimentally measured walks from the theoretical expectation. These effects include a different choice of the phase angle $\chi$ in Eq.~\eqref{eq-matrix} and the specific form of the ratchet initial state in Eq.~\eqref{eq-rat-2}, both reflecting the fact that we are dealing with an AOKR quantum walk. Note that residual peaks at low momenta observed in the AOKR quantum walk would not appear in an ideal quantum walk.

\begin{figure*}[tb]
    \centering
    \includegraphics[width=\linewidth]{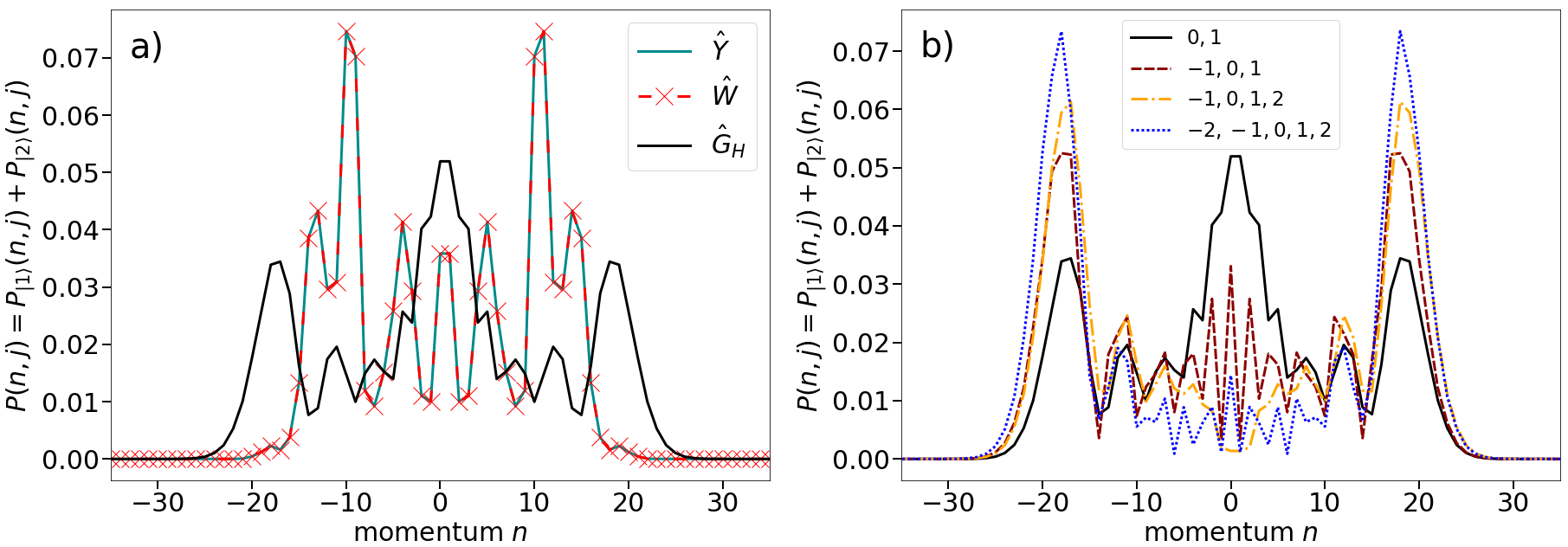}
    \caption{The walker's distributions are shown after $j=20$ steps for an AOKR discrete-time quantum walk. The kick strength is set at the experimental value $k=1.45$. In (a) different walk protocols are shown. The distributions are computed by evolving the initial state in momentum space given by Eq.~\eqref{eq-rat-2}. The label $\hat{Y}$ resembles the walk that is initialized by $\hat{W}$ and the evolution is executed with the $\hat{Y}$-coin. $\hat{W}$ and $\hat{G}_H$ are initialized by the $\hat{Y}$-coin and then their respective walk is executed by $\hat{W}$ or $\hat{G}_{\rm H}$. $\hat{Y}$ and $\hat{W}$ produce the same momentum distributions for all times.
    The AOKR walks in (b) are implemented by the $\hat{Y}$-coin and executed by the $\hat{G}_{\rm H}$-coin.  The different labels denote the momentum classes included in the initial state, as denoted by $s$ in Eq.~\eqref{eqratchet}. The broader the initial state is in momentum space the more the peak in the central region vanishes. One should remember that only the state expressed by Eq.~\eqref{eq-rat-2} (solid black line) was experimentally implemented in Refs.~\cite{Dadras2018, dadras2019experimental,Clark2021}. }
    \label{fig1}
\end{figure*}

\section{Novel Theoretical Model}

\subsection{Theory of the Light Shift}
\label{sec-physical_explanation}

The physical explanation is based on the additional light shift that starts playing a role in the spinor AOKR, described in detail in Refs.~\cite{Groiseau2017, Groiseau2018}. For clarity of the argument, we shall briefly present its origin here. 

During a kick, the dynamics of the standard AOKR are described by interaction terms between the ground and excited state of the form
\begin{equation}
        \expandafter\hat H_{\rm int} = \frac{\Omega}{2}|g\rangle\langle e|\cos(\expandafter\frac{\hat \theta}{2})e^{i\Delta t}+h.c.,
        \label{eq.effective_Hamiltonian}
\end{equation}
where $\Delta$ is the detuning and $\Omega$ the Rabi frequency of the laser.

The effective dynamics obtained after adiabatically eliminating the excited state are described by an AC-Stark shift of the ground states $|g\rangle$ from the coherent drive of the kicking laser, i.e., 
\begin{equation}
        \expandafter\hat H_{\rm eff} = \frac{\Omega^2}{8\Delta}|g\rangle\langle g|(\mathrm{cos}(\expandafter\hat \theta)+1),
        \label{eq.effective_Hamiltonian}
\end{equation}
where we used $\cos^2\frac{\theta}{2}=\frac{1}{2}(\cos\theta+1)$ and the rate corresponds to the kick strength before the time integration over the duration of the kick pulse $\tau_p$, e.g. $k=\frac{\Omega^2}{8\Delta}\tau_p$.

In the standard AOKR this constant offset (term with no $\cos\theta$-dependence) can be disregarded as there is only a single level. In our spinor AOKR, after the adiabatic elimination of the excited state, two ground states remain, each with such an AC-Stark shift (of opposite sign due to opposite detuning). Transitions between the two ground states can get discarded in rotating wave approximation. Thus, we are left with
\begin{equation}
        \expandafter\hat H_{\rm eff} = \frac{\Omega^2}{8\Delta}\hat \sigma_z(\mathrm{cos}(\expandafter\hat \theta)+1),
        \label{eq.effective_Hamiltonian}
\end{equation}
and effectively we have an additional energy difference or light shift between the two ground states which can no longer be discarded.

\subsection{Light-shift Compensation in the Experiment}
\label{sec-lightshift}

As just introduced and shown in full detail in Refs.~\cite{Groiseau2017, Groiseau2018}, the Hamiltonian for an AOKR with two different internal states contains an additional constant AC Stark shift \cite{Delone1999} between the two energy levels. Comparing the physically effectively implemented Hamiltonian from Eq.~\eqref{eq.effective_Hamiltonian} with the QKR-Hamiltonian from Eq.~\eqref{eq1}, this light shift induces a phase whenever a kick is applied giving an effective kick of the form
\begin{equation}
\hat{\mathcal{U}}_{\rm k, eff} = e^{-i \hat \sigma_z k \left( 1+ \cos(\hat{\theta}) \right)}.
\label{eq-lights}
\end{equation}
This means that there is a relative phase of $2k$ for each application of the kick operator, i.e., for each step of the walk. This light-shift phase needs to be compensated in the experiment since it would lead to a different evolution with respect to the theoretical prediction (note that the new terms in Eq.~\eqref{eq-lights} would adversely affect the phase evolution in the internal degree of freedom changing the overall interference pattern). A compensation with a $\hat \sigma_z$ phase gate with a third Bloch angle $\gamma=k$ by an additional MW pulse would be possible. The experiments reported in Refs.~\cite{Dadras2018, dadras2019experimental, Clark2021}, however, used the phase $\chi$ of Eq.~\eqref{eq-matrix} as a free parameter in order to best compensate the light shift phase. Several runs were made for various choices of $\chi$ and finally the value, with which the walk was most symmetric around $n=0$, was used in all other experiments in Refs.~\cite{dadras2019experimental,Dadras2018,Clark2021}. The absolute value of $\chi$ as well as a possibly present third Bloch angle $\gamma$ were
under limited experimental control, and the aforementioned compensation procedure seemed to make this fact irrelevant. 

The experiments may have, for instance, easily
exchanged the coin $\hat{Y}$ by the coin $\hat{G_H}$ in the walk, effectively resulting in a new sequence, e.g.,
\begin{equation}
    \hat{\mathcal{U}}_\mathrm{step}^j = [\hat{\mathcal{U}}\hat{G}_{\rm H}]^j \hat{\mathcal{U}} \hat{Y},
\label{eq:new sequence}
\end{equation}
The $\hat{Y} $ and $\hat{G}_H $ curves in Fig.~\ref{fig1}(a) show that such an exchange of the two coins indeed has dramatic effects on the quality of the walk. The operator $ \hat{G_{H}} $ is the Hadamard gate defined as~\cite{dadras2019experimental}:
\begin{equation}
\hat{G}_{\rm H} = \frac{1}{\sqrt{2}} \begin{pmatrix}
1 & 1 \\
1 & -1
\end{pmatrix} .
\label{eq5}
\end{equation}
While momentum distributions of the QWs represented by Eq.~\eqref{eq:new sequence} and Eq.~\eqref{eq-walk} are mirror symmetric around $n=0$ since both coins are perfectly balanced (all giving unbiased walks), the actual final distributions look very different. Assuming that only a MW pulse expressed by Eq.~\eqref{eq-matrix} was applied as stated in Refs.~\cite{Dadras2018, dadras2019experimental, Clark2021} with $\alpha=\pi/2$ fixed, the combined effect of a MW pulse and the light shift could have been of the form
\begin{eqnarray}
\hat{M}\left(\frac{\pi}{2},\chi\right)e^{-ik\hat\sigma_z }=
\frac{1}{\sqrt{2}}
\begin{pmatrix}
& e^{-ik} & e^{-i(\chi+k)}\\
& -e^{i(\chi-k)} & e^{ik}
\end{pmatrix}  \\
= \frac{e^{-ik}}{\sqrt{2}} \begin{pmatrix}
& 1 & e^{-i(\chi-2k)}\\
& -e^{i\chi} & e^{i2k}
\end{pmatrix} .
\label{eqlightshift}
\end{eqnarray}
In the last step we extracted a global phase $e^{-ik}$ that is not important for the following discussion. Generally, the phase $\chi$ cannot fully remove the effect of the light shift phase here. The quantum walk can, however, still be made symmetric around $n=0$ by the choice $\chi=2k=\pi \mod(2\pi)$, which would yield an effective MW operation.
Hence, the aforementioned swapping of the two different coin operators could have occurred in the experiments. For example, with a kick strength of $k\approx 1.5$ the light shift phase gives a value close to 
$2k\approx\pi$ (see Eq.~\eqref{eq-coin2}). Small deviations from the condition for $1.2 < k < 1.8$ appear not to change the global picture, as will be later shown in more detail in Sect.~\ref{sec:comparison}. In that sense, the light-shift and its experimentally incomplete compensation is the physical reasoning for the potentially implemented sequence from Eq.~\eqref{eq:new sequence}. 

\subsection{Alternative MW Pulse -- Hadamard Gate}
\label{sec-gate}

We have just seen that the actually implemented MW pulses in the experiment may be close to Hadamard gates $\hat{G}_{\rm H}$. %\cite{Nielsen}. 
In contrast to the original $\hat{W}$ pulses, $\hat{G}_{\rm H}$ pulses have the minus sign on the diagonal. Both pulses, however, are completely unbiased leading to walks with sidepeaks moving symmetrically outwards in a ballistic manner. We find that the difference in the signs of $ \hat{W} $ and $ \hat{G_{H}} $ matrix elements has no consequence for an ideal quantum walk with just nearest-neighbor couplings. For our AOKR walks, however, the different sign induces significantly different behavior. Figure~\ref{fig1}(a) shows a numerical example derived for a perfectly resonant walk ($\beta=0$). Our simulation results for the here proposed QW (see Eq.~\eqref{eq:new sequence}) clearly indicate that the bulk of its momentum distributions has a larger probability to remain in the center ($ n=0 $), as shown by the black curve in Fig.~\ref{fig1}(a). 

\subsection{Initial-state Dependence}
\label{sec-initial}

As described previously, an important difference between an ideal quantum walk and the AOKR walks discussed in this paper are the initial states in the walker's space \cite{Dadras2018}. The initial state experimentally implemented was expressed by Eq.~\eqref{eq-rat-2} with two involved momenta. As described in  Refs.~\cite{Ni2016, Ni2017, Mark2007, Gil2008}, the state is constructed to be concentrated in position space at the rising (falling) flanks of the potential where the force impulse towards the left (right) is maximal. It is exactly this effect that leads to directed ratchet-like motion. The more momentum states that are included in the initial state, the more densely peaked is the wavefunction in position (angle) space. For a highly dense wavefunction in position space, the directed motion works with minimal dispersion.
This dispersion is a specific problem in our AOKR walk with respect to an ideal quantum walk. Hence, it is indeed not too surprising that the AOKR QWs become more similar to ideal QWs when using "better" ratchet initial states. This is seen in Fig.~\ref{fig1}(b) for the walk with the new Hadamard coin $\hat G_{\rm H}$ during the evolution steps. The artificial clumping at the center of the momentum distributions disappears when more momentum classes are included in the initial states (see Fig.~\ref{fig1}(b)).

It is known that an ideal quantum walk does not display a central peak from the start, independent of the initial state (see Ref.~\cite{Kempe2003}). 
    The consequence is that an ideal walk does not display any difference between the various implementations using the different balanced coins described above. In the end, the dominant central peak, displayed when  using the $\hat{G}_{\rm H}$ coin, can be seen as an artifact from AOKR realization when using the simplest initial state. This central peak disappears when adding more momentum classes to the initial state (see Fig.~\ref{fig1}(b)). This provides a clear prediction that could easily be checked in future experiments.

In other words, the experimentally observed residual central peak is actually a relic of the AOKR dynamics. This behavior is expected when in the walk protocol due to light shift effects, the effectively implemented coin during the walk is $\hat{G}_H$ and not $\hat{Y}$, as initially intended. Even when this is the case, the central peak is only visible for an initial ratchet state sufficiently narrow in momentum space.

\subsection{Analytic Solution}
\label{sec-analytical}

\begin{figure}[b]
    \centering
    \includegraphics[width=0.9\linewidth]{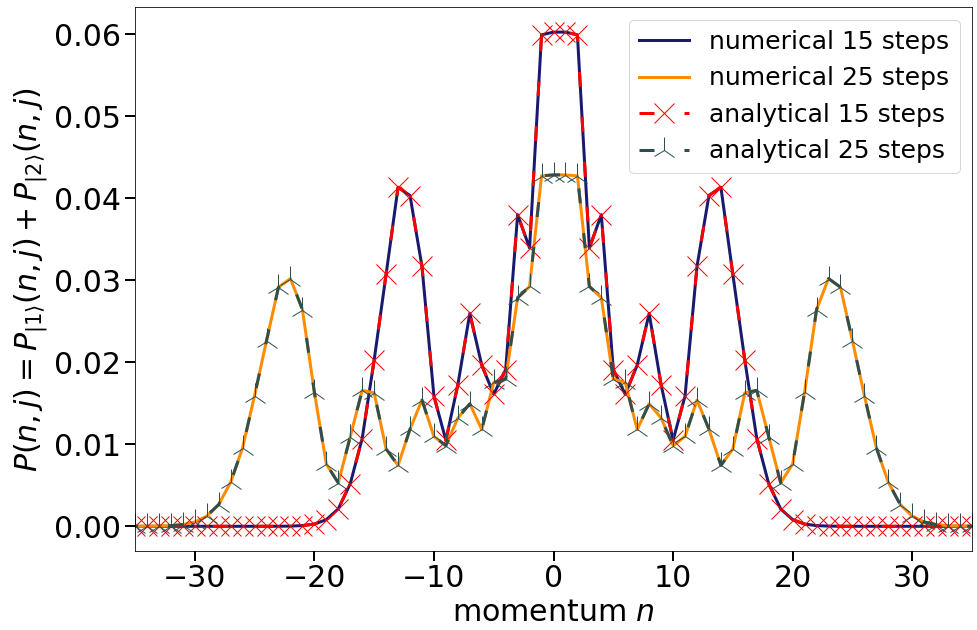}
    \caption{Comparison between numerical implementation of the walk and its analytical solution, as derived from Eq.~\eqref{eqmomentum}. As exemplary cases we show the final momentum distributions for $j=15$ and $j=25$ steps, with a kick strength $k=1.45$. The initial state in momentum space is given by Eq.~\eqref{eq-rat-2}. }
    \label{fig2}
\end{figure}

A comparison between the numerical implementation of the novel walk given by Eq.~\eqref{eq:new sequence} and the corresponding analytical solution derived from Eq.~\eqref{eqmomentum} is shown in Fig.~\ref{fig2}. The full calculation for the analytical expression is somewhat lengthy and reveals little insight as it closely follows Refs.~\cite{Groiseau2017, Groiseau2018}. Therefore, we only present here the final result for the momentum distributions, while the calculation in full detail can be found in the appendix. The final momentum distribution is 
\begin{widetext}
\begin{eqnarray}
\begin{split}
    P(n,j) & =  P_{|1\rangle}(n,j) + P_{|2\rangle}(n,j) 
    = \frac{1}{2^{j+1}S} \left[ \left(\sum_{l=0}^N \sum_s a_{l,1}(-1)^sJ_{(n-s)}\left ((N-2l-1)k \right) \right)^2 \right. \\
   &+ \left.\left(\sum_{l=0}^N \sum_s a_{l,2}(-1)^sJ_{(n-s)}\left ((N-2l+1)k \right) \right)^2 \right. 
    + \left. \left(\sum_{l=0}^N \sum_s a_{l,1}(-1)^sJ_{(n-s)}\left (-(N-2l-1)k \right) \right)^2 \right. \\ 
    &+ \left. \left(\sum_{l=0}^N \sum_s a_{l,2}(-1)^s J_{(n-s)}\left (-(N-2l+1)k \right) \right)^2 \right]
\end{split}
\label{eqmomentum}
\end{eqnarray}
%\allowdisplaybreaks
Here $J_\alpha(x)$ are Bessel functions of the first kind and 
the coefficients $a_{l,1/2}$ are given by %\begin{scriptsize}
\begin{eqnarray}
\begin{split}
      a_{l,1} &=
    \frac{1}{2^N} \sum_{u=0}^{N/2}  \sum_{m=0}^{l} \left( \binom{N}{2u} - \binom{N}{2u+1} \right) \binom{u}{m} \binom{N-2m}{l-m} (-1)^{N-l+m} \ 8^m \ \\\
     &+\frac{1}{2^N} 2 \sum_{u=0}^{N/2} \sum_{m=0}^{l-1}  \binom{N}{2u+1} \binom{u}{m} \binom{N-2m-1}{l-m-1} (-1)^{N-l+m} \ 8^m  \\\
     &-\frac{1}{2^N} 2 \sum_{u=0}^{N/2} \sum_{m=0}^{l}   \binom{N}{2u+1} \binom{u}{m} \binom{N-2m-1}{l-m} (-1)^{N-l+m} \ 8^m 
\end{split}
\label{eqa1}
\end{eqnarray}
and
\begin{equation}
a_{l,2}=  \frac{1}{2^{N}} \sum_{u=0}^{N/2} \sum_{m=0}^{l}  \binom{N+1}{2u+1} \binom{u}{m} \binom{N-2m}{l-m} (-1)^{-l+m}8^m \,,
\label{eqa2}
\end{equation}
\end{widetext}
%\end{scriptsize}
with ${N \equiv j-1}$. The sum over $s$ in Eq.~\eqref{eqmomentum} denotes the sum over the involved momentum classes in the initial state given by Eq.~\eqref{eqratchet}. Note that the momentum distribution is found to be of the same analytical form as those discussed in Ref.~\cite{Groiseau2018}. The coefficients only differ from previous results by a factor $(-1)^{-l}$ within the sums. These additional factors change the interference patterns in such a way that the different walk protocols, as discussed in Sect.~\ref{sec-gate}, lead to different momentum distributions. Since the result above is valid for an arbitrary number of walk steps, we arrived at a full understanding of the two different QWs with the two coins $\hat Y$ and $\hat W$ (or rather $\hat{G}_{\rm H}$) interchanged.

\subsection{Comparison between Theoretical Explanations}
\label{sec:comparison}

We have put forward an alternative way of understanding the central peaks around zero momentum in the experimental implementations of the AOKR quantum walks. To simulate experimental systems, we must include the finite width in the initial quasi-momentum distribution of the spinor BECs mentioned in Sect.~\ref{sec-intro}. This is best done numerically by averaging over a reasonable ensemble of quasi-momenta $\beta$ ~\cite{Groiseau2018}. Nonresonant $\beta$ induces a phase scrambling \cite{WGF2003, SW2011}, making the walks less ballistic with the effect of reducing the population in the ballistically moving sidepeaks. The value of $\beta$, drawn from a Gaussian distribution of a certain width $\beta_{\rm FWHM}$, was estimated in the experiments as $\beta_{\rm FWHM}\approx 0.025$ (see Refs.~\cite{Dadras2018, dadras2019experimental}). The numerical walks are obtained as an average over 1000 realizations, with each realization involving a value of $\beta$ being randomly drawn from the corresponding Gaussian.

\begin{figure}[htb]
	\includegraphics[width=\linewidth]{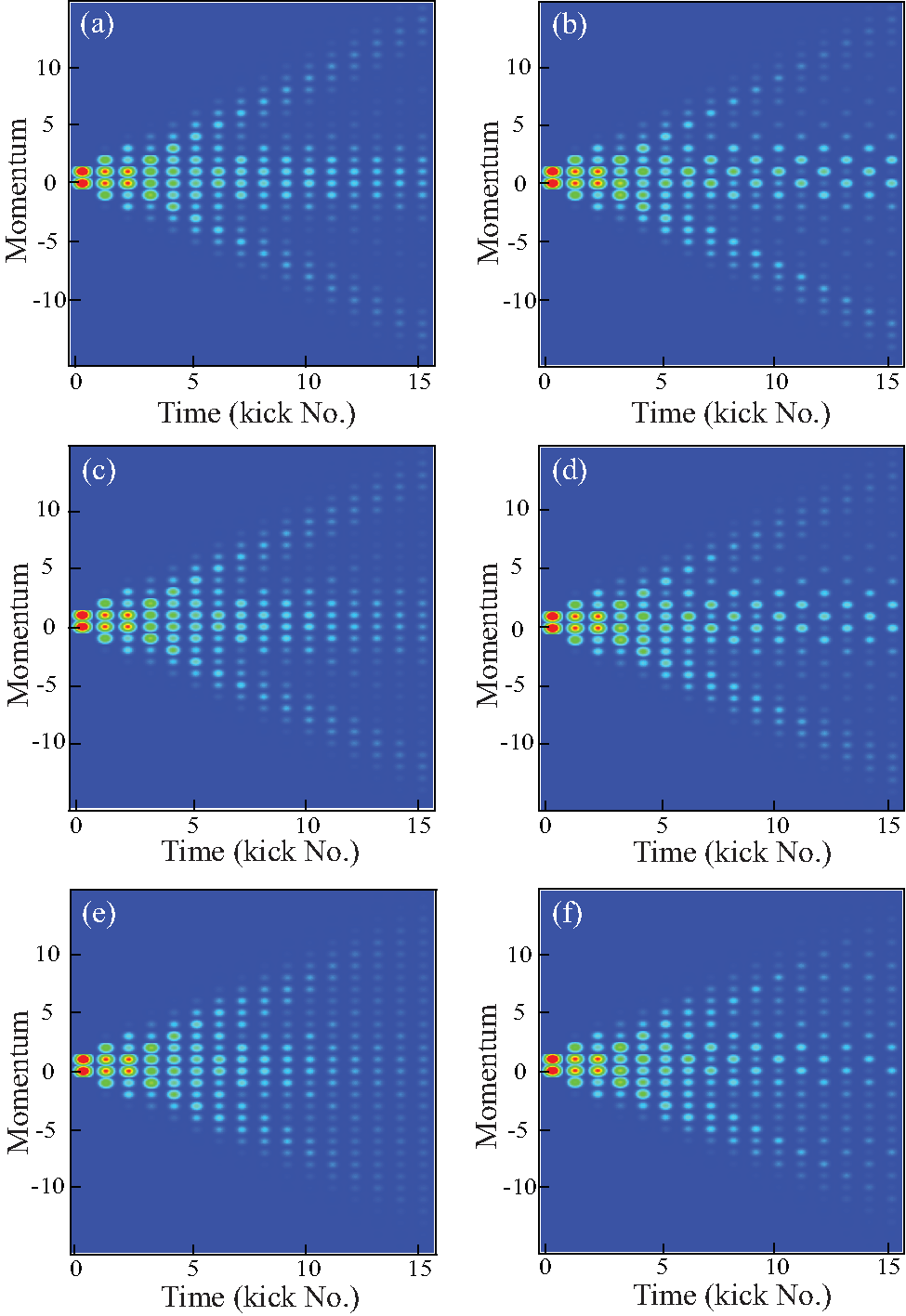}
    \caption{Numerical simulations of AOKR quantum walks with $k=1.45$ and different quasi-momentum distributions with $\beta_{\rm FWHM}=0$ (a,b), $\beta_{\rm FWHM}=0.01$ (c,d), and $\beta_{\rm FWHM}=0.025$ (e,f), all averaged over 1000 values of $\beta$. Left panels: implemented with the $\hat G_{\rm H}$ coin. Right panels: executed with Eq.~\eqref{eqlightshift} at $\chi=\pi$. 
    It can be seen that despite the small deviations, as discussed in Sect.~\ref{sec-lightshift}, both protocols essentially follow the same behavior,
    making both likely to correspond to the actual experimental data.}
    \label{fig3}  
\end{figure}

\begin{figure}[tb]
  \centering
%  \begin{tabular}{@{}c@{}}
    \includegraphics[width=.65\linewidth]{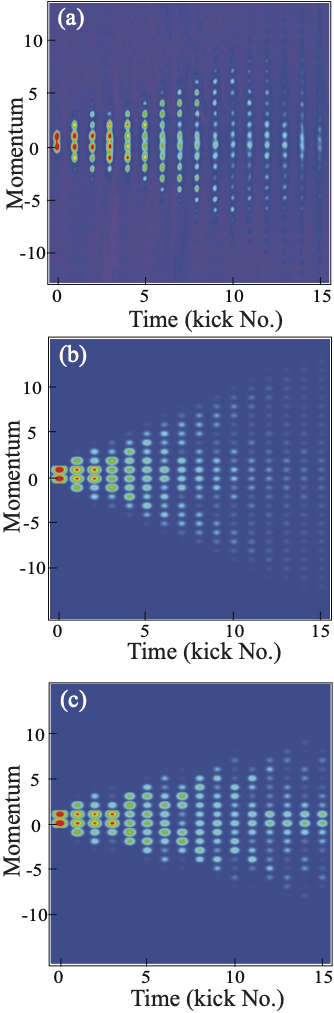}
    %\\[\abovecaptionskip]
%  \end{tabular}
%  \vspace{\floatsep}
%  \begin{tabular}{@{}c@{}}
  %  \includegraphics[width=0.92\linewidth]{fig5.jpg} %\\[\abovecaptionskip]
%  \end{tabular}
  \caption{
  AOKR quantum walks with $k=1.45$. (a) Experimental data adapted
  from Ref.~\cite{Dadras2018}, (b) Numerical simulation derived from our current theoretical model (see Eq.~\eqref{eq:new sequence}) with $\beta_{\rm FWHM}=0.025$ and using initial $ \hat{Y} $ rotation and $ \hat{G_{H}} $ coins. (c) Numerical simulation derived from the previous theoretical model (see Eq.~\eqref{eq-walk}) by adding a cloud of thermal atoms to the BEC part (see text).}
  \label{fig4}
\end{figure}

In the left panels of Fig.~\ref{fig3}, the walks are implemented by the $\hat{G}_{\rm H}$ coin, while the right panels feature the implementation of Eq.~(\ref{eqlightshift}). In other words, while the left panels show the novel walk that we argue to be responsible for the experimentally observed momentum distributions, the right panels show theoretical predictions using experimental parameters based on the originally proposed $\hat W$ coin and an incorrectly chosen compensation phase (see Eq.~\eqref{eqlightshift})
with $\chi=\pi$ and $k=1.45$. As anticipated in Sect. \ref{sec-lightshift}, the latter two protocols given by Eq.~\eqref{eq:new sequence} and Eq.~\eqref{eqlightshift} essentially lead to the same momentum distributions for all choices of $\beta_{\rm FWHM}=0$ in Fig.~\ref{fig3}(a,b), $\beta_{\rm FWHM}=0.01$ in Fig.~\ref{fig3}(c,d), and $\beta_{\rm FWHM}=0.025$ in Fig.~\ref{fig3}(e,f).
 With increasing $\beta_{\rm FWHM}$, the sidepeaks and the central regions become less and less distinct and the ballistic sidepeaks tend to fade out.  

Similar behavior is seen in our experimental data \cite{Dadras2018, dadras2019experimental, Clark2021}. Figure~\ref{fig4}(a) shows a typical experimental result adapted from Ref.~\cite{Dadras2018}. We find good theory-experiment agreements by comparing Fig.~\ref{fig4}(a) with Fig.~\ref{fig4}(b) that shows the predictions of our current model (see Eq.~\eqref{eq:new sequence}).
First, we observe in both Fig.~\ref{fig4}(a) and Fig.~\ref{fig4}(b) a central part that does not evolve far away from the origin and the two sidepeaks that evolve ballistically away from their initial position in momentum space. Second, the observed and predicted rates of the spread of these sidepeaks in momentum space with increasing number of steps appear comparable. Our current interpretation shown in Fig.~\ref{fig4}(b) would also be in reasonable agreement with the originally guessed temperature of the BEC with $\beta_{\rm FWHM}\approx 0.025$, when the fading of the sidepeaks is considered. Fig.~\ref{fig4}(c) shows momentum distributions of the QW given by the previous theoretical model (see Eq.~\eqref{eq-walk}) after a residual thermal cloud of atoms is added into the BECs. 

The thermal cloud was originally assumed as a possible solution for the appearance of the prominent central region. Thermal atoms essentially will not follow the kicking evolution \cite{WGF2003, SW2011} and hence remain close to the center. The experimentally intended $\hat{Y}$-protocol does not display this behavior, as can be seen from Fig.~\ref{fig1}. However, the QWs shown in Fig.~\ref{fig4}(c) appear to be different from our experimental observations, i.e., the predicted QWs lack the significantly contributing central region and the structure of the sidepeaks are of a quite different shape.

\begin{figure}[tb]
    \includegraphics[width=\linewidth]{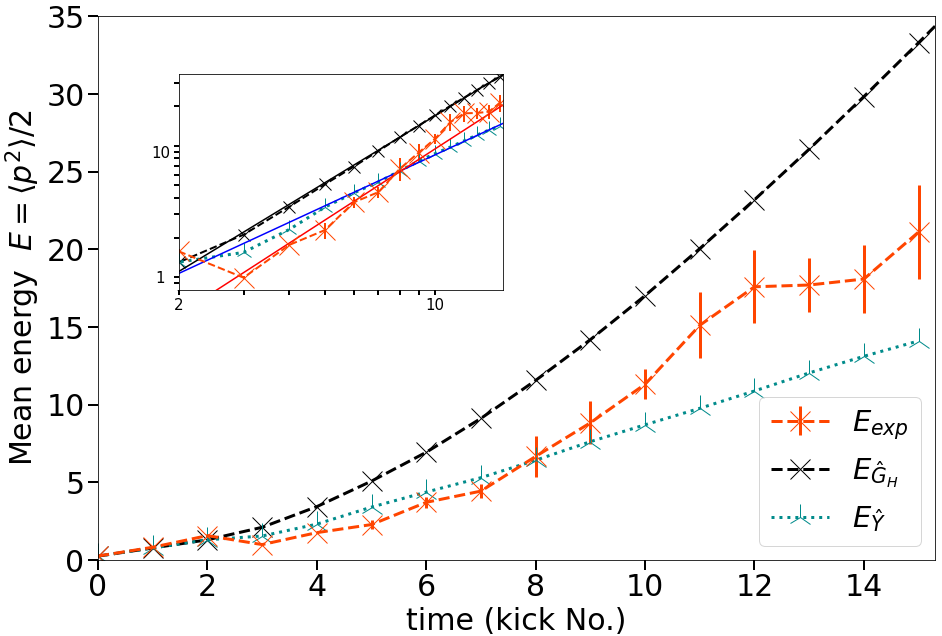}
    \caption{
    Comparison between the mean kinetic energies calculated from the walks seen in Fig.~\ref{fig4} with $\beta_{\rm FWHM}=0.025$.
    The energy, $E_{\rm exp}$, is extracted from the experimental walk data from Fig. \ref{fig4}(a). $E_{\hat{G}_{\rm H}}$ and $E_{\hat{Y}}$ 
    denote the mean energy for the walk executed by the $\hat{G}_H$-coin
    (Fig. \ref{fig4}(b)) and the $\hat{Y}$-coin (Fig. \ref{fig4}(c)), respectively. The inset shows the energies on a double logarithmic scale with power-law exponents 
    extracted by the fits (solid lines), giving $1.8\pm0.2$ for the experimental data and
    $1.7\pm0.1$ ($E_{\hat{G}_{\rm H}}$) and $1.3\pm0.1$ ($E_{\hat{Y}}$) for both theoretical models. The apparent better agreement between the fits for $E_{\rm exp}$ and $E_{\hat{G}_{\rm H}}$ confirms the better scaling of the novel model. The asymptotic exponent of 2 expected for a ballistic walk is hardly reached for quantum walks with just 15 steps.
    }   
   \label{fig5}
\end{figure}

The mean energies of the novel and original theoretical models were calculated and plotted as a function of time in Fig.~\ref{fig5} for comparison. It can be seen that the energy using $\hat{G}_{H}$-coin increases faster than that of the previous model. The increase in mean energy for the $\hat{Y}$-coin has a linear form, while the $\hat{G}_{H}$-coin increases more quadratically. Note that quantum resonant AOKR-walks possess a quadratic increase in mean energy, corresponding to a ballistic motion in momentum space. In the presence of a strong off-resonant $\beta$-distribution like the residual thermal cloud from the original theory, the energy increases only linearly \cite{WGF2003, SW2011}. The data shown for a small number of up to 15 steps maximum show that the asymptotic regimes are rarely met.

The mean energy extracted from experimental data presented in Fig.~\ref{fig4} (a) is also plotted in Fig.~\ref{fig5}. 
The experimentally obtained energy increases with more quadratic than linear behavior, which is better consistent with current theory and contradicts the presence of a thermal cloud as originally hypothesized. The comparison is yet more complicated since the experimental suffered from a series of well-known issues, see Ref. \cite{dadras2019experimental}. The effect most relevant in our context is the fading out of the ballistic peaks in the experimental momentum distributions due to atom number fluctuations and small atom losses. Each time slice is obtained from a new experimental run, and hence also the relative normalization of the atomic density, might be an issue. All this may have consequences on the second moment of the distribution that is proportional to the energy plotted in Fig.~\ref{fig5}. Counting less in the tails of the distribution typically leads to an underestimation of the mean energy \cite{WGF2003}.

In App. \ref{app-comp}, we also show further comparisons between our current theoretical model (see Eq.~\eqref{eq:new sequence}) and the previous experimental data, in particular similar plots as in Fig. \ref{fig4} for other values of the kick strength and a more direct matching of the momentum distributions for a specific case.

\section{Conclusion}
\label{sec:concl}

We have introduced a more complete theoretical explanation for the peculiar behavior observed in the discrete-time quantum walks implemented with the AOKR platform in Refs.~\cite{Dadras2018, dadras2019experimental, Clark2021}. We argue that the coin operations acting on the internal states of the atoms may have been different from the original proposal discussed in \cite{SW2016}. This difference, induced by the experimental calibration of the coin parameters together with an additional AC Stark shift present in the setup, may have led to less efficient quantum walks with a large population remaining close to the starting site of the walker. Our hypothesis may be checked in future experiments by either controlling much better the MW phases at compensated light shift or using ratchet states with less dispersion \cite{Ni2017} as initial states for the walks.

The understanding of the experimental results is of importance for further applications of walks realized with the AOKR platform. Our analysis implies that the realized walks may have had a higher quality than expected in the subsequent manner:
the central population seemingly not participating in the walker's evolution is actually an artificial interference effect induced by a non-optimal coin and therefore an ingredient of the system itself.
This effect makes the AOKR platform look even better for the quantum simulation of non-trivial walks and the investigation of applications such as quantum search algorithms \cite{Delvecchio2020} or of topological phases in Floquet-driven systems \cite{Groiseau2019}.

\acknowledgements
A. G. acknowledges support by the Deutsche Forschungsgemeinschaft (German Research Foundation), Project No. 441423094. J.H.C. and Y.L. thank the Noble Foundation for financial support.

\appendix

\onecolumngrid

\section{Details of the analytic calculation of the momentum distribution}

We consider one step of the novel quantum walk of interest and this is expressed by the operator $U$. The walk will be initialized by the $Y$-matrix.
\begin{equation}
 {U} \equiv { G}_{\rm H}=
\frac{1}{\sqrt{2}} \left( \begin{array}{cc}
1 &1 \\
1 & -1
\end{array} \right)
\left( \begin{array}{cc}
 e^{-ik \mathrm{cos}({\theta})    } & 0 \\
0 &  e^{ik \mathrm{cos}({\theta})    }
\end{array} \right)
\label{eq1}
\end{equation}
\begin{equation}
   {Y}  = \frac{1}{\sqrt{2}} \left( \begin{array}{cc}
1 &i \\
i & 1
\end{array} \right)
\end{equation}
To derive a closed form solution to the evolution we rewrite $U^T$ with ${N \equiv T-1} $ and ${ N \geq 0}$. We define $U^T$ in the following fashion:
\begin{equation}
 U^T = \left(\frac{1}{\sqrt{2}}\right)^T
\left( \begin{array}{rr}
A_1^{(T-1)}(k)  &  A_2^{(T-1)}(k) \\ 
A_3^{(T-1)}(k) &  A_4^{(T-1)}(k)   \end{array} \right) .
\label{eq2}
\end{equation}
Repeating the steps of the calculation in Ref.~\cite{Groiseau2018}, the goal is to express the matrix elements as polynomials in the kick operator $e^{\pm ik\cos(\mathrm{\theta})}$ and then translate the resulting evolution back to momentum space.
% by using \cite{abramowitz1964handbook} 
% \begin{equation}
%     {\int_0^{2\pi} e^{in\theta}e^{ik\mathrm{cos}\theta} d\theta = 2\pi i^n J_n(k)}.
%     \label{eq8}
% \end{equation}
Looking at the first few orders of the evolution, we notice that the matrix elements are related to each other according to
\begin{equation} 
 A_1^{(N)} (k) = (-1)^{(N+1)} A_4^{(N)}(-k) ,
\label{eq3}
\end{equation} \vspace{-0.8cm}
\begin{equation} 
 A_2^{(N)} (k) = (-1)^{(N+1)} A_3^{(N)}(-k) .
\label{eq4}
\end{equation}
The matrix entries will be found to be recursive polynomials. The initial conditions are

\begin{align}
p_1^{(0)}(\tilde{z}) = p_2^{(0)}(\tilde{z}) = 1,
 \label{eq5}
\end{align}\vspace{-0.8cm}
\begin{align}
p_1^{(1)}(\tilde{z}) = z \equiv e^{-ik\mathrm{cos}(\theta)}+e^{ik\mathrm{cos}(\theta)},
 \label{eq6}
\end{align}\vspace{-0.8cm}
\begin{align}
p_2^{(1)}(\tilde{z}) = \tilde{z} \equiv e^{-ik\mathrm{cos}(\theta)}-e^{ik\mathrm{cos}(\theta)}.
 \label{eq7}
\end{align}
We will show by induction that the following equations for calculating the matrix elements hold for arbitrary $N$:
\begin{align}
A_1^{(N)} (\tilde{z}) = e^{-ik\mathrm{cos}(\theta)}p_1^{(N)}(\tilde{z})
 \label{eq8}\\\
A_2^{(N)} (\tilde{z}) = e^{ ik\mathrm{cos}(\theta)}p_2^{(N)}(\tilde{z}),
 \label{eq9}
\end{align}
where the polynomials $p_1^{(N)}$ and $p_2^{(N)}$ are defined by a recursion formula.
\begin{equation}
p^{(N)} = \tilde{z}p^{(N-1)}+2p^{(N-2)}.
 \label{eq10}
\end{equation}
It will be demonstrated that the solution to the recursion can be written as polynomials in the kick operator, i.e.,
\begin{equation}
p_{1/2}^{(N)}= \sum_{l=0}^N a_{l,1/2} \ e^{ik \mathrm{cos}(\theta)(N-2l)}.
 \label{eq11}
\end{equation}
Comparing to Ref.~\cite{Groiseau2018}, one should notice slight differences. The computations will be somewhat analogous, but will also show differences in the details. In this section, the matrix elements in Eq.~\eqref{eq3} and Eq.~\eqref{eq4} have different relations amongst each other, compared to Ref.~\cite{Groiseau2018}.
Also the polynomials in Eq.~\eqref{eq5}-\eqref{eq7} are now functions in $\tilde{z}$ instead of having $z$ as their argument. Slight deviations are also found in Eq.~\eqref{eq8}-\eqref{eq9} and the recursion formula shows the same structure but varies from the one reported in Ref.~\cite{Groiseau2018} for the walk with $Y$ instead of $\hat G_{\rm H}$.

\section{Solution to the recursion formula}

Let us assume for now that Eq.~(\ref{eq10}) is true.
To solve the equation one may choose the ansatz $p^{(N)}(z) \equiv x^N(z)$ and plug it in which yields
\begin{equation}
    x^N = \tilde{z}x^{(N-1)}+2x^{(N-2)}.
    \label{eq12}
\end{equation}
This leads to a quadratic formula
\begin{equation}
    x^2 = \tilde{z}x+2,
    \label{eq13}
\end{equation}
with the solution
\begin{equation}
    x_{1/2} = \frac{\tilde{z} \pm \sqrt{\tilde{z}^2+8}}{2}.
    \label{eq14}
\end{equation}
The recursion formula from above satisfies linearity. Therefore the general solution is given by a linear combination of both solutions.
\begin{equation}
    p_{1/2}^{(N)}(z)=c_1x_1^{(N)}+c_2x_2^{(N)}
    \label{eq15}
\end{equation} 
The coefficients $c_1$ and $c_2$ can be found by putting in the initial conditions which leads to
\begin{equation}
    p_1^{(N)} = \frac{1}{2} \left( 1+\frac{2z-\tilde{z}}{\sqrt{\tilde{z}^2+8}} \right)\left( \frac{\tilde{z} + \sqrt{\tilde{z}^2+8}}{2}\right)^N+\frac{1}{2} \left( 1-\frac{2z-\tilde{z}}{\sqrt{\tilde{z}^2+8}} \right)\left( \frac{\tilde{z} - \sqrt{\tilde{z}^2+8}}{2}\right)^N,
    \label{eq16}
\end{equation}
and
\begin{equation}
    p_2^{(N)} = \frac{1}{2} \left( 1+\frac{\tilde{z}}{\sqrt{\tilde{z}^2+8}} \right)\left( \frac{\tilde{z} + \sqrt{\tilde{z}^2+8}}{2}\right)^N+\frac{1}{2} \left( 1-\frac{\tilde{z}}{\sqrt{\tilde{z}^2+8}} \right)\left( \frac{\tilde{z} - \sqrt{\tilde{z}^2+8}}{2}\right)^N.
    \label{eq17}
\end{equation}
Since the initial conditions and the recursion showed some differences from Ref.~\cite{Groiseau2018}, we find a solution for the polynomials that differs in several signs and as stated above, the polynomials have their argument in $\tilde{z}$.

\section{Prerequisites}
Before continuing some simple relations will be shown that will later prove themselves to be useful:
\begin{align}
\begin{split}
    z^2-\tilde{z}^2
    &\ =( e^{-ik\mathrm{cos}(\theta)}+e^{ik\mathrm{cos}(\theta)})^2-(e^{-ik\mathrm{cos}(\theta)}-e^{ik\mathrm{cos}(\theta)})^2 \\\
    &\ =e^{-2ik\mathrm{cos}(\theta)}+e^{2ik\mathrm{cos}(\theta)}+2-(e^{-2ik\mathrm{cos}(\theta)}+e^{2ik\mathrm{cos}(\theta)}-2)\\\
    &\ = 4
\end{split}\\[10pt]
\begin{split}
    (z-\tilde{z})(z\tilde{z}+\tilde{z}^2+8)
    &\ =\left(e^{-ik\mathrm{cos}(\theta)}+e^{ik\mathrm{cos}(\theta)}-(e^{-ik\mathrm{cos}(\theta)}-e^{ik\mathrm{cos}(\theta)}) \right)\\\
    &\ \cdot\left( (e^{-ik\mathrm{cos}(\theta)}+e^{ik\mathrm{cos}(\theta)})(e^{-ik\mathrm{cos}(\theta)}-e^{ik\mathrm{cos}(\theta)})+(e^{-ik\mathrm{cos}(\theta)}-e^{ik\mathrm{cos}(\theta)})^2+8 \right)\\\
    &\ =2e^{ik\mathrm{cos}(\theta)} \left(e^{-2ik\mathrm{cos}(\theta)}-e^{2ik\mathrm{cos}(\theta)}+(e^{-2ik\mathrm{cos}(\theta)}+e^{2ik\mathrm{cos}(\theta)}-2) +8\right)\\\
    &\ =2e^{ik\mathrm{cos}(\theta)}\left( 2e^{-2ik\mathrm{cos}(\theta)}+6\right) \hspace*{-0.1cm}=4\left(e^{-ik\mathrm{cos}(\theta)} +3e^{ik\mathrm{cos}(\theta)}\right)\\\
    &\ =4 \left(2(e^{-ik\mathrm{cos}(\theta)}+e^{ik\mathrm{cos}(\theta)})-2(e^{-ik\mathrm{cos}(\theta)}-e^{ik\mathrm{cos}(\theta)}) \right)\\\
    &\ =4(2z-\tilde{z})
\end{split}\\[10pt]
\begin{split}
   \tilde{z}^\alpha 
   &\ = ( e^{-ik\mathrm{cos}(\theta)}-e^{ik\mathrm{cos}(\theta)})^\alpha\\\
   &\ =\sum_{u=0}^\alpha\binom{\alpha}{u}(e^{-ik\mathrm{cos}(\theta)})^u(-e^{ik\mathrm{cos}(\theta)})^{\alpha-u}\\\
   &\ =\sum_{u=0}^\alpha \binom{\alpha}{u} (-1)^{\alpha-u} e^{ik\mathrm{cos}(\theta) (\alpha-2u)}
\end{split} \\[10pt]
\int_0^{2\pi} e^{in\theta}e^{ik\mathrm{cos}\theta} d\theta
&\ = 2\pi i^n J_n(k)\label{A21}\\[10pt]
J_{-a}(k)
&\ = (-1)^aJ_a(k)
\end{align}
Eq. \eqref{A21} is taken from Ref.~\cite{abramowitz1964handbook}.

\section{Proof for recursion}
We have already solved the recursion above by the ansatz. We still have to show that the solution of the recursion also solves for the matrix elements as stated in Eqs. (\ref{eq8}) and (\ref{eq9}).
The proof will follow by induction.
The statement is trivially true for $N=0$, by the choice of the initial conditions. Now let the statement be true for $N$. Then we will show that the statement will also be true for $N+1$. It is
\begin{equation} U^{(T+1)} \propto 
\left( \begin{array}{cc}
A_1^{(T-1)}(k)  &  A_2^{(T-1)}(k) \\ 
A_3^{(T-1)}(k) &  A_4^{(T-1)}(k)   \end{array} \right) 
 \left( \begin{array}{cc}
 e^{-ik \mathrm{cos}(\theta)    } & e^{ik \mathrm{cos}(\theta)}  \\
 e^{-ik \mathrm{cos}(\theta)    } & - e^{ik \mathrm{cos}(\theta)} 
\end{array} \right).\end{equation}
From that we can conclude:
\begin{align} 
A_1^{(N+1)} &= e^{-ik\mathrm{cos}(\theta)}(A_1^{(N)}+A_2^{(N)})\\
A_2^{(N+1)} &= e^{ik\mathrm{cos}(\theta)} (A_1^{(N)}-A_2^{(N)}).
\end{align}
First we proof the induction for $A_1$.

\begin{align}
    A_1^{(N+1)} &= e^{-ik\mathrm{cos}(\theta)} (A_1^{(N)}+A_2^{(N)})\\
    &= e^{-ik\mathrm{cos}(\theta)} \left( e^{-ik\mathrm{cos}(\theta)} p_1^{(N)} + e^{ik\mathrm{cos}(\theta)} p_2^{(N)} \right)\\
    &= e^{-ik\mathrm{cos}(\theta)} \left( \frac{z+\tilde{z}}{2}p_1^{(N)}+\frac{z-\tilde{z}}{2}p_2^{(N)} \right)\\
    &= e^{-ik\mathrm{cos}(\theta)}\left( \tilde{z} p_1^{(N)} + \frac{z-\tilde{z}}{2}(p_1^{(N)}+p_2^{(N)})\right)\\
    \begin{split}
        &=e^{-ik\mathrm{cos}(\theta)}
        \left[\tilde{z}p_1^{(N)} + \frac{z-\tilde{z}}{2} \left( \frac{1}{2} \left( 1+\frac{2z-\tilde{z}}{\sqrt{\tilde{z}^2+8}} \right)\left( \frac{\tilde{z} + \sqrt{\tilde{z}^2+8}}{2}\right)^N+\frac{1}{2} \left( 1-\frac{2z-\tilde{z}}{\sqrt{\tilde{z}^2+8}} \right)\left( \frac{\tilde{z} - \sqrt{\tilde{z}^2+8}}{2}\right)^N \right.\right.\\\
        &\qquad \qquad \qquad \left.\left. +   \frac{1}{2} \left( 1+\frac{\tilde{z}}{\sqrt{\tilde{z}^2+8}} \right)\left( \frac{\tilde{z} + \sqrt{\tilde{z}^2+8}}{2}\right)^N+\frac{1}{2} \left( 1-\frac{\tilde{z}}{\sqrt{\tilde{z}^2+8}} \right)\left( \frac{\tilde{z} - \sqrt{\tilde{z}^2+8}}{2}\right)^N \right) \right]
    \end{split}\\
    &= e^{-ik\mathrm{cos}(\theta)} \left[\tilde{z}p_1^{(N)} + \frac{z-\tilde{z}}{2} \left( \frac{1}{2} \left( 2+ \frac{2z}{\sqrt{\tilde{z}^2+8}}\right) \right)\left( \frac{\tilde{z} + \sqrt{\tilde{z}^2+8}}{2}\right)^N + \frac{1}{2} \left( 2- \frac{2z}{\sqrt{\tilde{z}^2+8}} \right)\left( \frac{\tilde{z} - \sqrt{\tilde{z}^2+8}}{2}\right)^N \right]\\
    \begin{split}
         &= e^{-ik\mathrm{cos}(\theta)}\left[ \tilde{z}p_1^{(N)} + \left( \left(\frac{z-\tilde{z}}{2} + \frac{z(z-\tilde{z})}{2\sqrt{\tilde{z}^2+8}} \right)\left(\frac{\tilde{z} + \sqrt{\tilde{z}^2+8}}{2}\right)^N\right.\right.  \\\
         &\qquad\qquad\qquad+ \left.\left( \frac{z-\tilde{z}}{2} - \left.+\frac{z(z-\tilde{z})}{2\sqrt{\tilde{z}^2+8}} \right) \left(\frac{\tilde{z} -\sqrt{\tilde{z}^2+8}}{2}\right)^N \right) \right] 
    \end{split}\\
    \begin{split}
        &= e^{-ik\mathrm{cos}(\theta)}
        \left[ \tilde{z}p_1^{(N)} + \left( \frac{(z-\tilde{z}\tilde{z})}{4} +\frac{z\tilde{z}(z-\tilde{z})}{4\sqrt{\tilde{z}+8}}+\frac{\sqrt{\tilde{z}+8}(z-\tilde{z})}{4} + \frac{z (z-\tilde{z})}{4} \right)\left(\frac{\tilde{z} + \sqrt{\tilde{z}^2+8}}{2}\right)^{N-1} \right.\\\
        & \qquad \qquad \qquad \left. + \left( \frac{(z-\tilde{z}\tilde{z})}{4} -\frac{z\tilde{z}(z-\tilde{z})}{4\sqrt{\tilde{z}+8}}-\frac{\sqrt{\tilde{z}+8}(z-\tilde{z})}{4} + \frac{z (z-\tilde{z})}{4} \right)\left(\frac{\tilde{z} - \sqrt{\tilde{z}^2+8}}{2}\right)^{N-1} \right]
    \end{split}\\
    \begin{split}
     &= e^{-ik\mathrm{cos}(\theta)}
     \left[\tilde{z}p_1^{(N)} + \left( \frac{(z-\tilde{z})(z\tilde{z}+\tilde{z}^2+8)}{4\sqrt{\tilde{z}^2+8}} + \frac{z^2-\tilde{z}^2}{4}\right)\left(\frac{\tilde{z} + \sqrt{\tilde{z}^2+8}}{2}\right)^{N-1} \right. \\\
    &\qquad \qquad \qquad \left.+ \left( -\frac{(z-\tilde{z})(z\tilde{z}+\tilde{z}^2+8)}{4\sqrt{\tilde{z}^2+8}} + \frac{z^2-\tilde{z}^2}{4}\right)\left(\frac{\tilde{z} - \sqrt{\tilde{z}^2+8}}{2}\right)^{N-1} \right]
    \end{split}\\
    &= e^{-ik\mathrm{cos}(\theta)}\left[\tilde{z}p_1^{(N)} + \left(1+\frac{2z-\tilde{z}}{\sqrt{\tilde{z}^2+8}}\right)\left(\frac{\tilde{z} + \sqrt{\tilde{z}^2+8}}{2}\right)^{N-1} + \left(1-\frac{2z-\tilde{z}}{\sqrt{\tilde{z}^2+8}}\right)\left(\frac{\tilde{z} - \sqrt{\tilde{z}^2+8}}{2}\right)^{N-1}   \right]\\
    &= e^{-ik\mathrm{cos}(\theta)}\left[ \tilde{z}p_1^{(N)} + 2p_1^{(N-1)} \right]
\end{align}
Therefore the induction for $A_1$ is complete. It remains to finish the induction for $A_2$.

\begin{align}
A_2^{(N+1)} &= e^{ik\mathrm{cos}(\theta)} (A_1^{(N)}-A_2^{(N)})\\
&= e^{ik\mathrm{cos}(\theta)}\left( e^{-ik\mathrm{cos}(\theta)}p_1^{(N)} - e^{ik\mathrm{cos}(\theta)}p_2^{(N)} \right)\\
&= e^{ik\mathrm{cos}(\theta)}\left( \frac{z+\tilde{z}}{2}p_1^{(N)} -\frac{z-\tilde{z}}{2}p_2^{(N)} \right)\\
&= e^{ik\mathrm{cos}(\theta)}\left( \tilde{z}p_2^{(N)} + \frac{z+\tilde{z}}{2}( p_1^{(N)} - p_2^{(N)} )\right)\\
\begin{split}
    &= e^{ik\mathrm{cos}(\theta)}
    \left[\tilde{z}p_2^{(N)} + \frac{z+\tilde{z}}{2}\left(\frac{1}{2} \left( 1+\frac{2z-\tilde{z}}{\sqrt{\tilde{z}^2+8}} \right)\left( \frac{\tilde{z} + \sqrt{\tilde{z}^2+8}}{2}\right)^N+\frac{1}{2} \left( 1-\frac{2z-\tilde{z}}{\sqrt{\tilde{z}^2+8}} \right)\left( \frac{\tilde{z} - \sqrt{\tilde{z}^2+8}}{2}\right)^N \right. \right.\\\
    & \qquad\qquad\qquad \left. \left. - \frac{1}{2} \left( 1+\frac{\tilde{z}}{\sqrt{\tilde{z}^2+8}} \right)\left( \frac{\tilde{z} + \sqrt{\tilde{z}^2+8}}{2}\right)^N-\frac{1}{2} \left( 1-\frac{\tilde{z}}{\sqrt{\tilde{z}^2+8}} \right)\left( \frac{\tilde{z} - \sqrt{\tilde{z}^2+8}}{2}\right)^N \right) \right]
\end{split}\\
&=e^{ik\mathrm{cos}(\theta)}\left[\tilde{z}p_2^{(N)} + \frac{z+\tilde{z}}{2} \left(\frac{1}{2} \frac{(z-\tilde{z})}{\sqrt{\tilde{z}^2+8}}\left( \frac{\tilde{z} + \sqrt{\tilde{z}^2+8}}{2}\right)^N +  \frac{1}{2} \frac{(-z+\tilde{z})}{\sqrt{\tilde{z}^2+8}}\left( \frac{\tilde{z} - \sqrt{\tilde{z}^2+8}}{2}\right)^N\ \right) \right]\\
&= e^{ik\mathrm{cos}(\theta)}\left[\tilde{z}p_2^{(N)} +  \frac{(z^2-\tilde{z}^2)}{2\sqrt{\tilde{z}^2+8}}  \left( \frac{\tilde{z} + \sqrt{\tilde{z}^2+8}}{2}\right)^N +\frac{(\tilde{z}^2-z^2)}{2\sqrt{\tilde{z}^2+8}}\left( \frac{\tilde{z} - \sqrt{\tilde{z}^2+8}}{2}\right)^N \right]\\
\begin{split}
    &= e^{ik\mathrm{cos}(\theta)}
    \left[\tilde{z}p_2^{(N)} + \left(\frac{\tilde{z}(z^2-\tilde{z}^2)}{4\sqrt{\tilde{z}^2+8}}+\frac{z^2-\tilde{z}^2}{4}  \right)\left( \frac{\tilde{z} + \sqrt{\tilde{z}^2+8}}{2}\right)^{N-1}\right.\\\
    &\qquad\qquad\qquad + \left.\left(\frac{\tilde{z}(z^2-\tilde{z}^2)}{4\sqrt{\tilde{z}^2+8}}-\frac{\tilde{z}^2-z^2}{4}  \right)\left( \frac{\tilde{z} - \sqrt{\tilde{z}^2+8}}{2}\right)^{N-1} \right]
\end{split}\\
&= e^{ik\mathrm{cos}(\theta)}\left[\tilde{z}p_2^{(N)} + \left(\frac{\tilde{z}}{\sqrt{\tilde{z}^2+8}}+1  \right)\left( \frac{\tilde{z} + \sqrt{\tilde{z}^2+8}}{2}\right)^{N-1} + \left(-\frac{\tilde{z}}{\sqrt{\tilde{z}^2+8}}+1  \right)\left( \frac{\tilde{z} - \sqrt{\tilde{z}^2+8}}{2}\right)^{N-1} \right]\\
&= e^{ik\mathrm{cos}(\theta)}\left[\tilde{z}p_2^{(N)} + 2p_2^{(N-1)} \right]
\end{align}
Therefore it is proven that $A_1$ and $A_2$ follow the relations given above. $A_3$ and $A_4$ can from here on be calculated by (\ref{eq3}) and (\ref{eq4}). 

\section{Rewriting the polynomials}

To obtain the final momentum distribution, it is convenient to rewrite the polynomials into a more accessible form. Therefore, the polynomials will be rewritten into polynomials in the kick operator 

\begin{align} p_1^{N} &= \frac{1}{2} \left( 1+\frac{2z-\tilde{z}}{\sqrt{\tilde{z}^2+8}} \right)\left( \frac{\tilde{z} + \sqrt{\tilde{z}^2+8}}{2}\right)^N+\frac{1}{2} \left( 1-\frac{2z-\tilde{z}}{\sqrt{\tilde{z}^2+8}} \right)\left( \frac{\tilde{z} - \sqrt{\tilde{z}^2+8}}{2}\right)^N\\
&= \frac{1}{2^{N+1}}\left[ \left(1+\frac{2z-\tilde{z}}{\sqrt{\tilde{z}^2+8}} \right) \left( \tilde{z} + \sqrt{\tilde{z}^2+8}\right)^N + \left( 1-\frac{2z-\tilde{z}}{\sqrt{\tilde{z}^2+8}} \right) \left( \tilde{z}-\sqrt{\tilde{z}^2+8}\right) ^N \right] \\
&= \frac{1}{2^{N+1}} \left[\left( \tilde{z}+\sqrt{\tilde{z}^2+8}\right) ^N + \left( \tilde{z}-\sqrt{\tilde{z}^2+8}\right) ^N + \frac{2z-\tilde{z}}{\sqrt{\tilde{z}^2+8}} \left( \left( \tilde{z}+\sqrt{\tilde{z}^2+8}\right) ^N - \left( \tilde{z}-\sqrt{\tilde{z}^2+8}\right) ^N \right) \right]\\
\begin{split}\label{A48}
    &= \frac{1}{2^{N+1}}
    \left[ \sum_{u=0}^N \binom{N}{u} \tilde{z}^{N-u} \left(\sqrt{\tilde{z}^2+8}\right)^u +  \sum_{u=0}^N \binom{N}{u} \tilde{z}^{N-u}\left(-\sqrt{\tilde{z}^2+8}\right)^u \right. \\\
    &\qquad\qquad \left. +\frac{2z-\tilde{z}}{\sqrt{\tilde{z}^2+8}} \left(\sum_{u=0}^N \binom{N}{u} \tilde{z}^{N-u}\left(\sqrt{\tilde{z}^2+8}\right)^u - \sum_{u=0}^N \binom{N}{u} \tilde{z}^{N-u}\left(-\sqrt{\tilde{z}^2+8}\right)^u \right) \right]
\end{split}\\
&=\frac{1}{2^N} \left[\sum_{u=0}^{N/2} \binom{N}{2u} \tilde{z}^{N-2u} \left(\tilde{z}^2+8\right)^u + \frac{2z-\tilde{z}}{\sqrt{\tilde{z}^2+8}} \left(\sum_{u=0}^{N/2} \binom{N}{2u+1} \tilde{z}^{N-2u-1} \left(\sqrt{\tilde{z}^2+8}\right)^{2u+1}\right) \right]\\
&=\frac{1}{2^N}\left[\sum_{u=0}^{N/2} \left( \binom{N}{2u} - \binom{N}{2u+1} \right) \tilde{z}^{N-2u}\left(\tilde{z}^2+8 \right)^u + 2 \left( \sum_{u=0}^{N/2} \binom{N}{2u+1}z \tilde{z}^{N-2u-1}\left(\tilde{z}^2+8\right)^u\right) \right] \\
&=\frac{1}{2^N}\left[\sum_{u=0}^{N/2} \sum_{m=0}^{u}  \left( \binom{N}{2u} - \binom{N}{2u+1} \right) \binom{u}{m} \tilde{z}^{N-2m}8^m +2\sum_{u=0}^{N/2} \sum_{m=0}^{u} \binom{N}{2u+1}\binom{u}{m}z \tilde{z}^{N-2m-1} 8^m \right]\\
\begin{split}
    &= \frac{1}{2^N} \left[\sum_{u=0}^{N/2}  \sum_{m=0}^{u} \sum_{l=0}^{N-2m} \left( \binom{N}{2u} - \binom{N}{2u+1} \right) \binom{u}{m} \binom{N-2m}{l} (-1)^{N-2m-l}\cdot8^m \cdot e^{ik \mathrm{cos}(\theta)(N-2m-2l)} \right] \\\
    &\quad +\frac{1}{2^N} 2 \left[\sum_{u=0}^{N/2} \sum_{m=0}^{u} \sum_{l=0}^{N-2m-1}  \binom{N}{2u+1} \binom{u}{m} \binom{N-2m-1}{l} (-1)^{N-2m-l-1}\cdot8^m \cdot e^{ik \mathrm{cos}(\theta)(N-2m-2l-2)}\right] \\\
    &\quad +\frac{1}{2^N} 2 \left[\sum_{u=0}^{N/2} \sum_{m=0}^{u} \sum_{l=0}^{N-2m-1}  \binom{N}{2u+1} \binom{u}{m} \binom{N-2m-1}{l} (-1)^{N-2m-l-1}\cdot 8^m  \cdot e^{ik \mathrm{cos}(\theta)(N-2m-2l)}\right] 
\end{split}\\
\begin{split}
    &= \frac{1}{2^N} \left[\sum_{u=0}^{N/2}  \sum_{m=0}^{u} \sum_{l=0}^{N-2m} \left( \binom{N}{2u} - \binom{N}{2u+1} \right) \binom{u}{m} \binom{N-2m}{l} (-1)^{N-l}\cdot8^m \cdot e^{ik \mathrm{cos}(\theta)(N-2m-2l)} \right] \\\
    &\quad -\frac{1}{2^N} 2 \left[\sum_{u=0}^{N/2} \sum_{m=0}^{u} \sum_{l=0}^{N-2m-1}  \binom{N}{2u+1} \binom{u}{m} \binom{N-2m-1}{l} (-1)^{N-l}\cdot8^m \cdot e^{ik \mathrm{cos}(\theta)(N-2m-2l-2)}\right] \\\
    &\quad -\frac{1}{2^N} 2 \left[\sum_{u=0}^{N/2} \sum_{m=0}^{u} \sum_{l=0}^{N-2m-1}  \binom{N}{2u+1} \binom{u}{m} \binom{N-2m-1}{l} (-1)^{N-l}\cdot 8^m  \cdot e^{ik \mathrm{cos}(\theta)(N-2m-2l)}\right] 
\end{split}\\
&= \sum_{l=0}^N a_{l,1} e^{ik \mathrm{cos}(\theta)(N-2l)}
\end{align}
where in the last step we have changed indices from $l$ to $l+m$.  Note that for the upper part in Eq. \eqref{A48} all odd powers cancelled each other, same account for the uneven powers of the lower part.
Furthermore, we defined $a_{l,1}$ to be

\begin{equation}
\begin{split}
    a_{l,1} =
    &\ \frac{1}{2^N} \sum_{u=0}^{N/2}  \sum_{m=0}^{l} \left( \binom{N}{2u} - \binom{N}{2u+1} \right) \binom{u}{m} \binom{N-2m}{l-m} (-1)^{N-l+m}\cdot8^m\\\
    &\ +\frac{1}{2^N} 2 \sum_{u=0}^{N/2} \sum_{m=0}^{l-1}  \binom{N}{2u+1} \binom{u}{m} \binom{N-2m-1}{l-m-1} (-1)^{N-l+m}\cdot 8^m  \\\
    &\ -\frac{1}{2^N} 2 \sum_{u=0}^{N/2} \sum_{m=0}^{l}   \binom{N}{2u+1} \binom{u}{m} \binom{N-2m-1}{l-m} (-1)^{N-l+m}\cdot 8^m  .
\end{split} 
\label{A55}
\end{equation}
Note the factor $(-1)$ dragged out in the third line of the coefficient $a_{l,1}$. This factor $(-1)$ has a different reason in Ref.~\cite{Groiseau2018} since it is not resolved from the shift of the index. Analogous steps have to be taken for $p_2^{(N)}$. 

\begin{align}
p_2^{(N)} &= \frac{1}{2} \left( 1+\frac{\tilde{z}}{\sqrt{\tilde{z}^2+8}} \right)\left( \frac{\tilde{z} + \sqrt{\tilde{z}^2+8}}{2}\right)^N+\frac{1}{2} \left( 1-\frac{\tilde{z}}{\sqrt{\tilde{z}^2+8}} \right)\left( \frac{\tilde{z} - \sqrt{\tilde{z}^2+8}}{2}\right)^N \\
& = \frac{1}{2^{N+1}} \left[\left( 1+\frac{\tilde{z}}{\sqrt{\tilde{z}^2+8}} \right) \left(\tilde{z} + \sqrt{\tilde{z}^2+8} \right)^N + \left( 1-\frac{\tilde{z}}{\sqrt{\tilde{z}^2+8}} \right) \left(\tilde{z} - \sqrt{\tilde{z}^2+8} \right)^N \right] \\
&= \frac{1}{2^{N+1}} \left[\left(\tilde{z} + \sqrt{\tilde{z}^2+8} \right)^N + \left(\tilde{z} - \sqrt{\tilde{z}^2+8} \right)^N + \frac{\tilde{z}}{\sqrt{\tilde{z}^2+8}}\left(\tilde{z} + \sqrt{\tilde{z}^2+8} \right)^N - \frac{\tilde{z}}{\sqrt{\tilde{z}^2+8}} \left(\tilde{z} - \sqrt{\tilde{z}^2+8} \right)^N \right] \\
\begin{split}
    &= \frac{1}{2^{N+1}}
     \left[\sum_{u=0}^N \binom{N}{u} \tilde{z}^{N-u} \left( \sqrt{\tilde{z}^2+8}\right)^u + \sum_{u=0}^N \binom{N}{u} \tilde{z}^{N-u} \left(- \sqrt{\tilde{z}^2+8}\right)^u\right. \\\
    &\qquad\qquad \left. +\frac{\tilde{z}}{\sqrt{\tilde{z}^2+8}} \left(\sum_{u=0}^N \binom{N}{u} \tilde{z}^{N-u} \left( \sqrt{\tilde{z}^2+8}\right)^u - \sum_{u=0}^N \binom{N}{u} \tilde{z}^{N-u} \left(- \sqrt{\tilde{z}^2+8}\right)^u \right)\right]
\end{split}\\
&= \frac{1}{2^{N}}\left[\sum_{u=0}^{N/2} \binom{N}{2u} \tilde{z}^{N-2u} \left( \tilde{z}^2+8 \right)^u + \frac{\tilde{z}}{\sqrt{\tilde{z}^2+8}} \sum_{u=0}^{N/2} \binom{N}{2u+1} \tilde{z}^{N-2u-1} \left(\sqrt{ \tilde{z}^2+8} \right)^{2u+1} \right]\\
&= \frac{1}{2^{N}}\left[\sum_{u=0}^{N/2} \binom{N}{2u} \tilde{z}^{N-2u} \left( \tilde{z}^2+8 \right)^u +  \sum_{u=0}^{N/2} \binom{N}{2u+1} \tilde{z}^{N-2u} \left( \tilde{z}^2+8 \right)^{u} \right] \\
& = \frac{1}{2^{N}}\left[\sum_{u=0}^{N/2} \binom{N+1}{2u+1} \tilde{z}^{N-2u} \left( \tilde{z}^2+8 \right)^u \right] \\
& = \frac{1}{2^{N}}\left[\sum_{u=0}^{N/2} \sum_{m=0}^{u} \binom{N+1}{2u+1} \binom{u}{m} \tilde{z}^{N-2m} 8^m \right] \\
& = \frac{1}{2^{N}}\left[\sum_{u=0}^{N/2} \sum_{m=0}^{u} \sum_{l=0}^{N-2m} \binom{N+1}{2u+1} \binom{u}{m} \binom{N-2m}{l} (-1)^{N-2m-l} \cdot 8^m \cdot e^{ik \mathrm{cos}(\theta) (N-2m-2l)}\right] \\
& = \frac{1}{2^{N}}\left[\sum_{u=0}^{N/2} \sum_{m=0}^{u} \sum_{l=0}^{N-2m} \binom{N+1}{2u+1} \binom{u}{m} \binom{N-2m}{l} (-1)^{N-l} \cdot 8^m \cdot e^{ik \mathrm{cos}(\theta) (N-2m-2l)}\right] \\&= \frac{1}{2^{N}} \sum_{l=0}^N a_{l,2} \cdot e^{ik \mathrm{cos}(\theta) (N-2l)}
\end{align}
Again in the last step we shifted the index $l$ shifted to $l+m$ and $a_{l,2} $ has been defined to be:

\begin{equation}a_{l,2} = \frac{1}{2^{N}} \sum_{u=0}^{N/2} \sum_{m=0}^{l}  \binom{N+1}{2u+1} \binom{u}{m} \binom{N-2m}{l-m} (-1)^{N-l+m} \cdot 8^m 
\label{A67}
\end{equation}

Please note that in Eqs. \eqref{A55} and \eqref{A67} the factor $(-1)^N$ acts like a global phase on the system and is therefore irrelevant. 

\section{Calculating the momentum-distribution }

The intention is to implement a ratchet state of the form:
\begin{equation}
    |\psi_{\rm R}\rangle = \frac{1}{\sqrt{S}}\sum_s e^{is\pi/2} |n=s\rangle.
    \label{eqratchet}
\end{equation}

As already stated above, the ratchet state is initialized by application of the beamsplitter matrix $\hat Y$. We organise our internal basis $|1\rangle \equiv \binom{0}{1}$ and $|2\rangle \equiv \binom{1}{0}$. Therefore, the total initial state is given by

\begin{equation}
    \psi_{in} =\frac{1}{\sqrt{2}}\left(i|1\rangle +|2\rangle \right) \otimes \frac{1}{\sqrt{S}}\sum_s e^{-is\frac{\pi}{2}} |n=s\rangle.
\end{equation}

The total momentum distribution is given as the sum of the momentum distributions of both the respective internal states:
{\small \begin{align} P(n,T) &= P_{|1\rangle}(n,T) + P_{|2\rangle}(n,T) \\
&=\left[ \left| \frac{1}{\sqrt{2\pi}}\int_0^{2\pi} e^{-in\theta} \langle\theta,1|U^T|\psi_{in}\rangle d\theta\right|^2 + \left| \frac{1}{\sqrt{2\pi}}\int_0^{2\pi} e^{-in\theta} \langle\theta,2|U^T|\psi_{in}\rangle d\theta\right|^2\right]\\
\begin{split}
    & =
    \left[ \left|  \frac{1}{\sqrt{2\pi}}\int_0^{2\pi} e^{-in\theta}\left(\frac{1}{\sqrt{2}}\right)^T \langle\theta,1 | {\left( \begin{array}{rr}
    A_1^{(T-1)}(k)  &  A_2^{(T-1)}(k) \\ 
    A_3^{(T-1)}(k) &  A_4^{(T-1)}(k)   \end{array} \right)}\frac{1}{\sqrt{2}}\left(i|1\rangle +|2\rangle \right) \otimes \frac{1}{\sqrt{S}}\sum_s e^{-is\frac{\pi}{2}} |n=s\rangle \right|^2 \right.\\\
    &\quad +\left. \left| \frac{1}{\sqrt{2\pi}}\int_0^{2\pi} e^{-in\theta}\left(\frac{1}{\sqrt{2}}\right)^T \langle\theta,2 | {\left( \begin{array}{rr}
    A_1^{(T-1)}(k)  &  A_2^{(T-1)}(k) \\ 
    A_3^{(T-1)}(k) &  A_4^{(T-1)}(k)   \end{array} \right)}\frac{1}{\sqrt{2}}\left(i|1\rangle +|2\rangle \right) \otimes \frac{1}{\sqrt{S}}\sum_s e^{-is\frac{\pi}{2}} |n=s\rangle \right|^2 \right] 
\end{split}\\
\begin{split}
    &=\frac{1}{2^{T+1}S}
      \left[ \left| \frac{1}{\sqrt{2\pi}}\int_0^{2\pi} e^{-in\theta} \left( iA_1^{(T-1)}+A_2^{(T-1)}\right) \sum_s (-i)^s \langle\theta|n=s\rangle d\theta \right|^2 \right. \\\
    &\qquad\qquad +\left. \left| \frac{1}{\sqrt{2\pi}}\int_0^{2\pi} e^{-in\theta} \left( iA_3^{(T-1)}+A_4^{(T-1)}\right) \sum_s (-i)^s \langle\theta|n=s\rangle d\theta \right|^2 \right]
\end{split}\\
\begin{split}
    &= \frac{1}{2^{T+1}S}
\left[ \left| \frac{1}{2\pi}\int_0^{2\pi} e^{-in\theta} \left( iA_1^{(T-1)}+A_2^{(T-1)}\right) \sum_s (-i)^s e^{is\theta} d\theta \right|^2 \right. \\\
&\qquad\qquad +\left. \left| \frac{1}{2\pi}\int_0^{2\pi} e^{-in\theta} \left( iA_3^{(T-1)}+A_4^{(T-1)}\right) \sum_s (-i)^s e^{is\theta} d\theta \right|^2 \right]
\end{split}\\
\begin{split}
    &=  \frac{1}{2^{T+1}S}\left[ \left| \frac{1}{2\pi}\int_0^{2\pi}\sum_s (-i)^s e^{-i(n-s)\theta} \left( i\sum_{l=0}^Na_{l,1}e^{ik\mathrm{cos}\theta(N-2l-1)} +\sum_{l=0}^Na_{l,2}e^{ik\mathrm{cos}\theta(N-2l+1)}\right)   d\theta \right|^2 \right. \\\
    &\qquad + \left. \left| \frac{1}{2\pi}\int_0^{2\pi}\sum_s (-i)^s e^{-i(n-s)\theta} \left( i\sum_{l=0}^N (-1)^N a_{l,1}e^{-ik\mathrm{cos}\theta(N-2l-1)} +\sum_{l=0}^N (-1)^{N+1}a_{l,2}e^{-ik\mathrm{cos}\theta(N-2l+1)}\right)   d\theta \right|^2 \right] 
\end{split}\\
\begin{split}
    &= \frac{1}{2^{T+1}S}
    \left[ \left| i\sum_{l=0}^N \sum_s a_{l,1}(-i)^s i^{-(n-s)}J_{-(n-s)}\left ((N-2l-1)k \right)\right.\right.\\\
    &\qquad\qquad\left. \left.+ \ \sum_{l=0}^N \sum_s a_{l,2}(-i)^s i^{-(n-s)}J_{-(n-s)}\left ((N-2l+1)k \right) \right|^2\right.\\\
    &\qquad\qquad +  \left. \left|i \sum_{l=0}^N \sum_s a_{l,1}(-1)^N(-i)^s i^{-(n-s)}J_{-(n-s)}\left (-(N-2l-1)k \right)\right.\right.\\\
    &\qquad\qquad \left.\left. + \ \sum_{l=0}^N \sum_s a_{l,2}(-1)^{N+1}(-i)^s i^{-(n-s)}J_{-(n-s)}\left (-(N-2l+1)k \right) \right|^2\right]
\end{split}\\
\begin{split}
    &= \frac{1}{2^{T+1}S} 
    \left[ \left| i\sum_{l=0}^N \sum_s a_{l,1} (-1)^{(n-s)}J_{(n-s)}\left ((N-2l-1)k \right) + \sum_{l=0}^N \sum_s a_{l,2}(-1)^{(n-s)}J_{(n-s)}\left ((N-2l+1)k \right) \right|^2\right.\\\
    &\qquad\qquad +  \left. \left|i \sum_{l=0}^N \sum_s a_{l,1}(-1)^{(n-s)}J_{(n-s)}\left (-(N-2l-1)k \right) + \sum_{l=0}^N \sum_s a_{l,2} (-1)^{(n-s)}J_{(n-s)}\left (-(N-2l+1)k \right) \right|^2\right]
\end{split}\\
\begin{split}\label{eqmomentum}
    &= \frac{1}{2^{T+1}S}
    \left[ \left(\sum_{l=0}^N \sum_s a_{l,1}(-1)^sJ_{(n-s)}\left ((N-2l-1)k \right) \right)^2 \right.
     + \left.\left(\sum_{l=0}^N \sum_s a_{l,2}(-1)^sJ_{(n-s)}\left ((N-2l+1)k \right) \right)^2 \right.+ \\\
     &\qquad\qquad \left. \left(\sum_{l=0}^N \sum_s a_{l,1}(-1)^sJ_{(n-s)}\left (-(N-2l-1)k \right) \right)^2 \right.
     + \left. \left(\sum_{l=0}^N \sum_s a_{l,2}(-1)^s J_{(n-s)}\left (-(N-2l+1)k \right) \right)^2 \right].
\end{split}
\end{align}
}
Even though the computation shows variations in its details, the momentum distributions found in Eq.~\eqref{eqmomentum} and in Ref.~\cite{Groiseau2018} are of the same analytical form. Nevertheless, the coefficients differ from each other.

\section{Comparison between the theoretical models and experimental data}
\label{app-comp}

In Refs.~\cite{Dadras2018, dadras2019experimental, Clark2021} the experiments are performed for several kick strengths, namely $k=1.2$, $k=1.45$, and $k=1.8$. In the previous section we worked out the analytical solution for resonant quasi-momentum $\beta=0$, valid for all parameters and initial states. Now, we include a finite distribution of quasi-momenta centered around zero with a finite width $\beta_{\rm FHWM}$ as described in the main text. This is done numerically as a closed-form for $\beta \neq 0$ is possible \cite{Groiseau2018} but can only be evaluated approximately.

\begin{figure}[tb]
    \centering
    \includegraphics[width=0.7\linewidth]{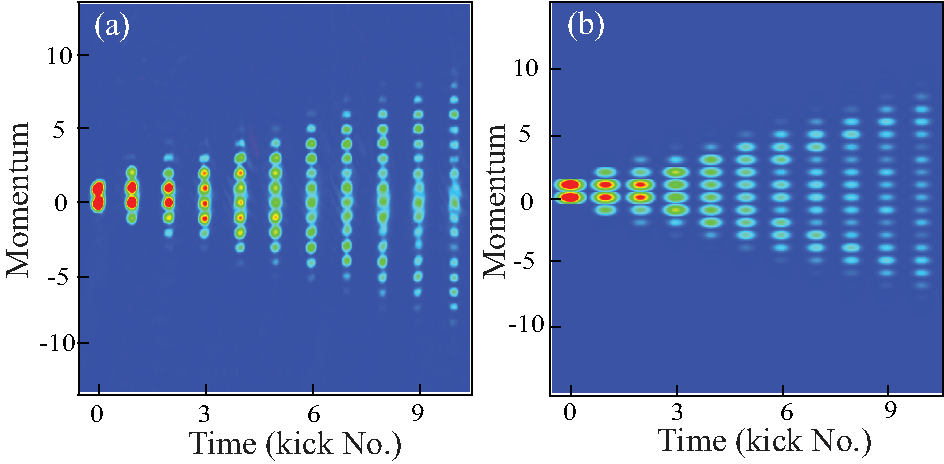}
    \caption{(Left) Experimentally realized quantum walks at $k=1.2$ adapted from Ref.~\cite{dadras2019experimental}. (Right) Our theoretical prediction derived from the proposed model given by Eq.(10) of the main text at $ k = 1.2 $}
    \label{fig:1}
\end{figure}

\begin{figure}[tb]
    \centering
    \includegraphics[width=0.7\linewidth]{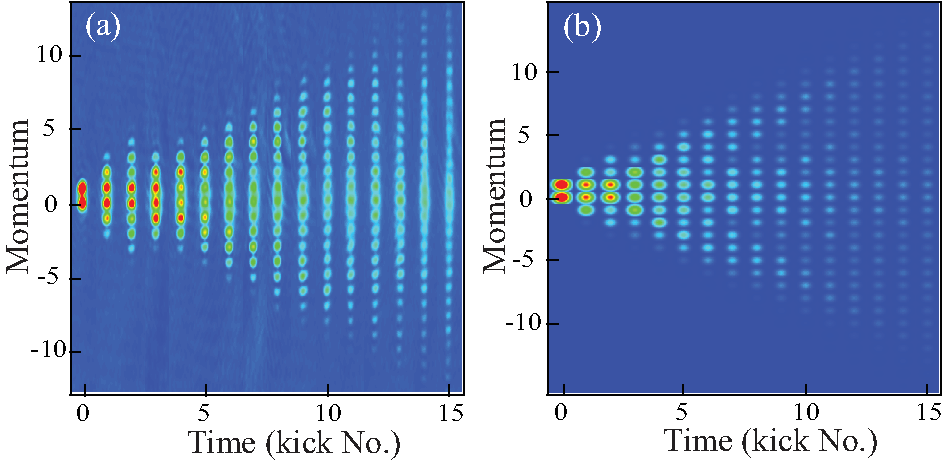}
    \caption{Same as in Fig. \ref{fig:1} for $k=1.45$. }
    \label{fig:2}
\end{figure}

\begin{figure}[tb]
    \centering
    \includegraphics[width=0.71\linewidth]{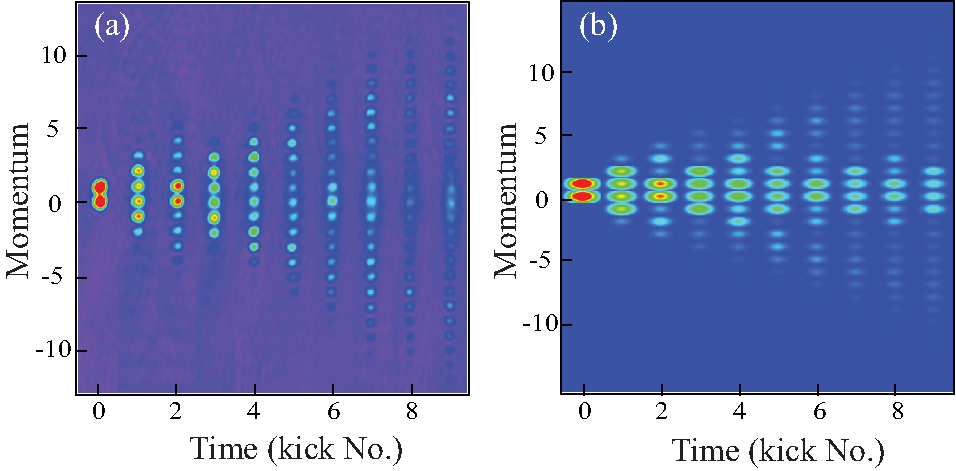}
    \caption{Same as in Fig. \ref{fig:1} for $k=1.8$.}
    \label{fig:3}
\end{figure}

We find a few interesting theory-experiment agreements as shown in Fig.~S\ref{fig:1}, Fig.~S\ref{fig:2}, and Fig.~S\ref{fig:3}. First, the theoretical results at several kick strengths $k$ feature the same characteristics as the experimental implementation, such as a dominant central peak and ballistically evolving sidearms. Second, for larger $k$, the sidearms appear to spread more with increasing steps of the walk. The sidearms in the experimental images tend to fade out for larger step numbers in the walks. This effect can be understood as a result from dephasing not only due to the nonresonant quasi-momenta but also due to other noise sources. Only for $k=1.8$ in Fig.~S\ref{fig:3}, however, the central peak is more dominant than expected by the experimental image. 

\renewcommand{\figurename}{Figure S}

%We noted that adding more nonresonant quasi-momenta with $\beta_{\rm FWHM} \geq 0.02$ enhances the fading out of the sidearms more than actually seen in the original experimental data. This indicates another important consequence of our new interpretation of the experiments of \cite{Dadras2018, dadras2019experimental, Clark2021}, namely that the condensate was colder than expected. We find now 
%the best comparison between theorey and experiment for $\beta_{\rm FWHM} \approx 0.01$ rather than the previous value $\beta_{\rm FWHM} \approx 0.025$ obtained from a similar comparison with the old theoretical interpretation including a relatively large thermal cloud of atoms.

\begin{figure}[t]
    \centering
    \includegraphics[width=0.8\linewidth]{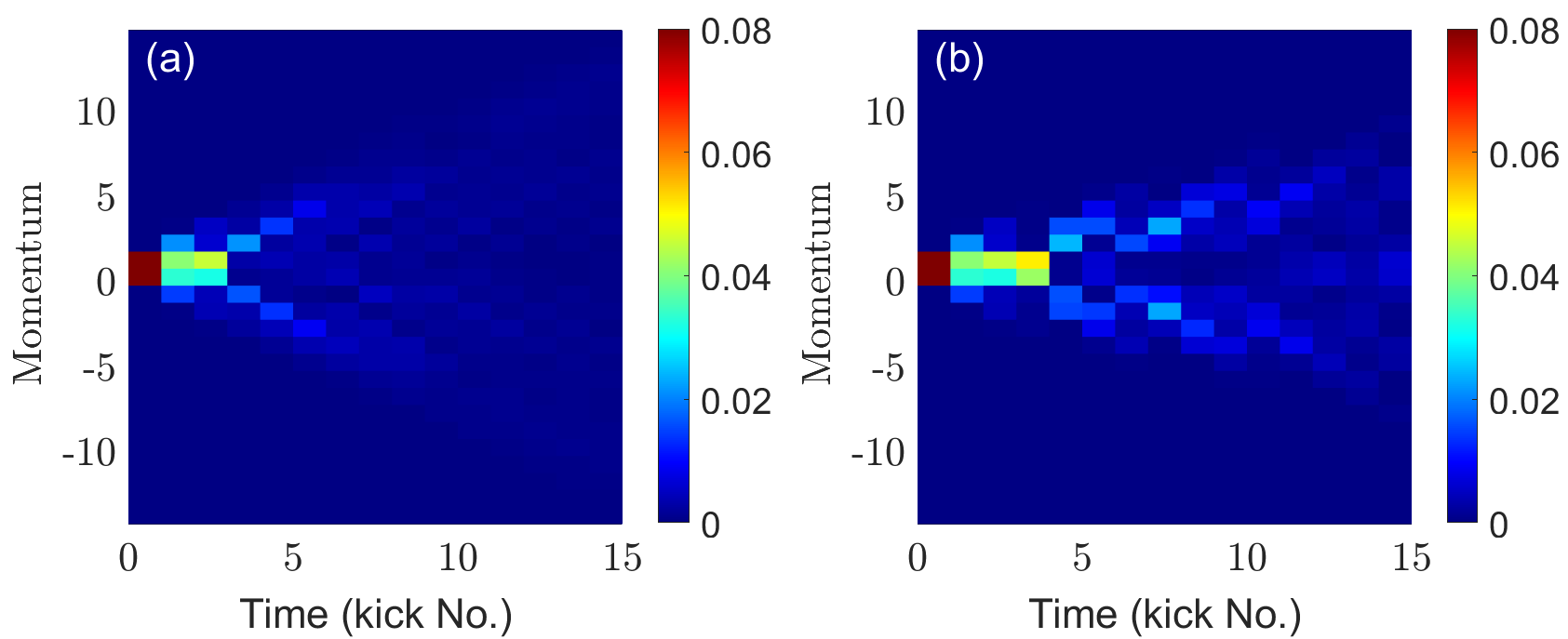}
    \caption{Comparison between experimental data and theoretical model by the difference values from Eq.~\eqref{eq.comparison} with $a=3$. Panel (a) shows comparison between the experiment and theoretical walk using the novel $\hat{G}_H$-coin. Panel (b) compares the experiment with a theoretical walk implementing the previous $\hat{Y}$-coin.}
    \label{fig:4}
\end{figure}

An in-depth direct comparison between the experimental data and the theoretical simulation of the proposed model is also attempted for walks with a kick strength of k=1.45. 
For this purpose, the information of the original experimental data set is reduced by comparing only the central value for each momentum class. This is necessary since the experimental images were taken in continuous momentum space, not in discrete integer valued momentum space as used in theoretical calculations. Thus as a consequence, the normalization of the full momentum distribution was not necessarily unity. 
As a result, the respective value of experimental data can be compared with both the original and present theoretical models for each momentum class at each step using the following expression:
\begin{equation}
    \mathrm{comparison}=\frac{|\mathrm{walk}_\mathrm{exp}-a \times \mathrm{walk}_\mathrm{theo}|}{a\times \mathrm{walk}_\mathrm{theo}}
    \label{eq.comparison}
\end{equation}
The experimental and theoretical walks are represented by "$\mathrm{walk}_\mathrm{exp}$" and "$\mathrm{walk}_\mathrm{theo}$" respectively.
Because the experimental data is not normalized, a rescaling factor $a$ is applied to the respective theoretical walk. Here, a scaling factor of $a=3$ is chosen because the extracted experimental distribution sums to about 3 for each time step, while the theoretical data is normalized.
An additional complication is the loss of probability over time in the experimental data from left to right. To account for this problem would necessitate introducing a time-dependent scaling factor to obtain a better comparison. Since in the latter the time-dependent fitting function is too arbitrary, we report the difference of Eq.~\eqref{eq.comparison} with a constant $a$.
Figure~S\ref{fig:4} presents the comparison according to Eq.~\eqref{eq.comparison} between experimental values and theoretical walks obtained using the coins $\hat{G}_H$ and $\hat{Y}$-coin, respectively. Better theory-experiment agreement can be seen in Fig.~S\ref{fig:4} (a), where the $\hat{G}_H$-coin is being proposed and corresponds to current theory. The total error, calculated from the total sum of the plotted pixels, is $0.88$ and $1.2$ in Fig.~S\ref{fig:4} (a) and Fig.~S\ref{fig:4} (b), respectively. Note in particular that the fading arms of the theoretical momentum distribution of the $\hat{Y}$-coin walk in Fig.~S\ref{fig:4} (b) are not reproduced in the experimental data. 
Although these aforementioned restrictions make comparison between previous experimental results with the theory challenging, we are nevertheless able to confirm in a quantitative manner that the current theory using the $\hat{G}_H$-coin provides better theory-experiment agreement regarding the shape of the walk than that of the previous theory using a $\hat{Y}$-coin.

\twocolumngrid


%%%\bibliography{citation.bib}

\begin{thebibliography}{37}%
\makeatletter
\providecommand \@ifxundefined [1]{%
 \@ifx{#1\undefined}
}%
\providecommand \@ifnum [1]{%
 \ifnum #1\expandafter \@firstoftwo
 \else \expandafter \@secondoftwo
 \fi
}%
\providecommand \@ifx [1]{%
 \ifx #1\expandafter \@firstoftwo
 \else \expandafter \@secondoftwo
 \fi
}%
\providecommand \natexlab [1]{#1}%
\providecommand \enquote  [1]{``#1''}%
\providecommand \bibnamefont  [1]{#1}%
\providecommand \bibfnamefont [1]{#1}%
\providecommand \citenamefont [1]{#1}%
\providecommand \href@noop [0]{\@secondoftwo}%
\providecommand \href [0]{\begingroup \@sanitize@url \@href}%
\providecommand \@href[1]{\@@startlink{#1}\@@href}%
\providecommand \@@href[1]{\endgroup#1\@@endlink}%
\providecommand \@sanitize@url [0]{\catcode `\\12\catcode `\$12\catcode
  `\&12\catcode `\#12\catcode `\^12\catcode `\_12\catcode `\%12\relax}%
\providecommand \@@startlink[1]{}%
\providecommand \@@endlink[0]{}%
\providecommand \url  [0]{\begingroup\@sanitize@url \@url }%
\providecommand \@url [1]{\endgroup\@href {#1}{\urlprefix }}%
\providecommand \urlprefix  [0]{URL }%
\providecommand \Eprint [0]{\href }%
\providecommand \doibase [0]{https://doi.org/}%
\providecommand \selectlanguage [0]{\@gobble}%
\providecommand \bibinfo  [0]{\@secondoftwo}%
\providecommand \bibfield  [0]{\@secondoftwo}%
\providecommand \translation [1]{[#1]}%
\providecommand \BibitemOpen [0]{}%
\providecommand \bibitemStop [0]{}%
\providecommand \bibitemNoStop [0]{.\EOS\space}%
\providecommand \EOS [0]{\spacefactor3000\relax}%
\providecommand \BibitemShut  [1]{\csname bibitem#1\endcsname}%
\let\auto@bib@innerbib\@empty
%</preamble>
\bibitem [{\citenamefont {Kempe}(2003)}]{Kempe2003}%
  \BibitemOpen
  \bibfield  {author} {\bibinfo {author} {\bibfnamefont {J.}~\bibnamefont
  {Kempe}},\ }\bibfield  {title} {\bibinfo {title} {Quantum random walks: An
  introductory overview},\ }\href@noop {} {\bibfield  {journal} {\bibinfo
  {journal} {Contemporary Physics}\ }\textbf {\bibinfo {volume} {44}},\
  \bibinfo {pages} {307} (\bibinfo {year} {2003})}\BibitemShut {NoStop}%
\bibitem [{\citenamefont {Portugal}(2018)}]{Portugal}%
  \BibitemOpen
  \bibfield  {author} {\bibinfo {author} {\bibfnamefont {R.}~\bibnamefont
  {Portugal}},\ }\href@noop {} {\emph {\bibinfo {title} {Quantum Walks and
  Search Algorithms}}}\ (\bibinfo  {publisher} {Springer International
  Publishing},\ \bibinfo {address} {New York},\ \bibinfo {year}
  {2018})\BibitemShut {NoStop}%
\bibitem [{\citenamefont {Lovett}\ \emph {et~al.}(2010)\citenamefont {Lovett},
  \citenamefont {Cooper}, \citenamefont {Everitt}, \citenamefont {Trevers},\
  and\ \citenamefont {Kendon}}]{Kendon}%
  \BibitemOpen
  \bibfield  {author} {\bibinfo {author} {\bibfnamefont {N.~B.}\ \bibnamefont
  {Lovett}}, \bibinfo {author} {\bibfnamefont {S.}~\bibnamefont {Cooper}},
  \bibinfo {author} {\bibfnamefont {M.}~\bibnamefont {Everitt}}, \bibinfo
  {author} {\bibfnamefont {M.}~\bibnamefont {Trevers}},\ and\ \bibinfo {author}
  {\bibfnamefont {V.}~\bibnamefont {Kendon}},\ }\bibfield  {title} {\bibinfo
  {title} {Universal quantum computation using the discrete-time quantum
  walk},\ }\href {https://doi.org/10.1103/PhysRevA.81.042330} {\bibfield
  {journal} {\bibinfo  {journal} {Phys. Rev. A}\ }\textbf {\bibinfo {volume}
  {81}},\ \bibinfo {pages} {042330} (\bibinfo {year} {2010})}\BibitemShut
  {NoStop}%
\bibitem [{\citenamefont {Dadras}\ \emph {et~al.}(2018)\citenamefont {Dadras},
  \citenamefont {Gresch}, \citenamefont {Groiseau}, \citenamefont {Wimberger},\
  and\ \citenamefont {Summy}}]{Dadras2018}%
  \BibitemOpen
  \bibfield  {author} {\bibinfo {author} {\bibfnamefont {S.}~\bibnamefont
  {Dadras}}, \bibinfo {author} {\bibfnamefont {A.}~\bibnamefont {Gresch}},
  \bibinfo {author} {\bibfnamefont {C.}~\bibnamefont {Groiseau}}, \bibinfo
  {author} {\bibfnamefont {S.}~\bibnamefont {Wimberger}},\ and\ \bibinfo
  {author} {\bibfnamefont {G.~S.}\ \bibnamefont {Summy}},\ }\bibfield  {title}
  {\bibinfo {title} {Quantum walk in momentum space with a {Bose-Einstein}
  condensate},\ }\href {https://doi.org/10.1103/PhysRevLett.121.070402}
  {\bibfield  {journal} {\bibinfo  {journal} {Phys. Rev. Lett.}\ }\textbf
  {\bibinfo {volume} {121}},\ \bibinfo {pages} {070402} (\bibinfo {year}
  {2018})}\BibitemShut {NoStop}%
\bibitem [{\citenamefont {Clark}\ \emph {et~al.}(2021)\citenamefont {Clark},
  \citenamefont {Groiseau}, \citenamefont {Shaw}, \citenamefont {Dadras},
  \citenamefont {Binegar}, \citenamefont {Wimberger}, \citenamefont {Summy},\
  and\ \citenamefont {Liu}}]{Clark2021}%
  \BibitemOpen
  \bibfield  {author} {\bibinfo {author} {\bibfnamefont {J.~H.}\ \bibnamefont
  {Clark}}, \bibinfo {author} {\bibfnamefont {C.}~\bibnamefont {Groiseau}},
  \bibinfo {author} {\bibfnamefont {Z.~N.}\ \bibnamefont {Shaw}}, \bibinfo
  {author} {\bibfnamefont {S.}~\bibnamefont {Dadras}}, \bibinfo {author}
  {\bibfnamefont {C.}~\bibnamefont {Binegar}}, \bibinfo {author} {\bibfnamefont
  {S.}~\bibnamefont {Wimberger}}, \bibinfo {author} {\bibfnamefont {G.~S.}\
  \bibnamefont {Summy}},\ and\ \bibinfo {author} {\bibfnamefont
  {Y.}~\bibnamefont {Liu}},\ }\bibfield  {title} {\bibinfo {title} {Quantum to
  classical walk transitions tuned by spontaneous emissions},\ }\href
  {https://doi.org/10.1103/PhysRevResearch.3.043062} {\bibfield  {journal}
  {\bibinfo  {journal} {Phys. Rev. Research}\ }\textbf {\bibinfo {volume}
  {3}},\ \bibinfo {pages} {043062} (\bibinfo {year} {2021})}\BibitemShut
  {NoStop}%
\bibitem [{\citenamefont {Dadras}\ \emph {et~al.}(2019)\citenamefont {Dadras},
  \citenamefont {Gresch}, \citenamefont {Groiseau}, \citenamefont {Wimberger},\
  and\ \citenamefont {Summy}}]{dadras2019experimental}%
  \BibitemOpen
  \bibfield  {author} {\bibinfo {author} {\bibfnamefont {S.}~\bibnamefont
  {Dadras}}, \bibinfo {author} {\bibfnamefont {A.}~\bibnamefont {Gresch}},
  \bibinfo {author} {\bibfnamefont {C.}~\bibnamefont {Groiseau}}, \bibinfo
  {author} {\bibfnamefont {S.}~\bibnamefont {Wimberger}},\ and\ \bibinfo
  {author} {\bibfnamefont {G.~S.}\ \bibnamefont {Summy}},\ }\bibfield  {title}
  {\bibinfo {title} {Experimental realization of a momentum-space quantum
  walk},\ }\href@noop {} {\bibfield  {journal} {\bibinfo  {journal} {Physical
  Review A}\ }\textbf {\bibinfo {volume} {99}},\ \bibinfo {pages} {043617}
  (\bibinfo {year} {2019})}\BibitemShut {NoStop}%
\bibitem [{\citenamefont {Preiss}\ \emph {et~al.}(2015)\citenamefont {Preiss},
  \citenamefont {Ma}, \citenamefont {Tai}, \citenamefont {Lukin}, \citenamefont
  {Rispoli}, \citenamefont {Zupancic}, \citenamefont {Lahini}, \citenamefont
  {Islam},\ and\ \citenamefont {Greiner}}]{preiss2015strongly}%
  \BibitemOpen
  \bibfield  {author} {\bibinfo {author} {\bibfnamefont {P.~M.}\ \bibnamefont
  {Preiss}}, \bibinfo {author} {\bibfnamefont {R.}~\bibnamefont {Ma}}, \bibinfo
  {author} {\bibfnamefont {M.~E.}\ \bibnamefont {Tai}}, \bibinfo {author}
  {\bibfnamefont {A.}~\bibnamefont {Lukin}}, \bibinfo {author} {\bibfnamefont
  {M.}~\bibnamefont {Rispoli}}, \bibinfo {author} {\bibfnamefont
  {P.}~\bibnamefont {Zupancic}}, \bibinfo {author} {\bibfnamefont
  {Y.}~\bibnamefont {Lahini}}, \bibinfo {author} {\bibfnamefont
  {R.}~\bibnamefont {Islam}},\ and\ \bibinfo {author} {\bibfnamefont
  {M.}~\bibnamefont {Greiner}},\ }\bibfield  {title} {\bibinfo {title}
  {Strongly correlated quantum walks in optical lattices},\ }\href@noop {}
  {\bibfield  {journal} {\bibinfo  {journal} {Science}\ }\textbf {\bibinfo
  {volume} {347}},\ \bibinfo {pages} {1229} (\bibinfo {year}
  {2015})}\BibitemShut {NoStop}%
\bibitem [{\citenamefont {D{\"u}r}\ \emph {et~al.}(2002)\citenamefont
  {D{\"u}r}, \citenamefont {Raussendorf}, \citenamefont {Kendon},\ and\
  \citenamefont {Briegel}}]{dur2002quantum}%
  \BibitemOpen
  \bibfield  {author} {\bibinfo {author} {\bibfnamefont {W.}~\bibnamefont
  {D{\"u}r}}, \bibinfo {author} {\bibfnamefont {R.}~\bibnamefont
  {Raussendorf}}, \bibinfo {author} {\bibfnamefont {V.~M.}\ \bibnamefont
  {Kendon}},\ and\ \bibinfo {author} {\bibfnamefont {H.-J.}\ \bibnamefont
  {Briegel}},\ }\bibfield  {title} {\bibinfo {title} {Quantum walks in optical
  lattices},\ }\href@noop {} {\bibfield  {journal} {\bibinfo  {journal}
  {Physical Review A}\ }\textbf {\bibinfo {volume} {66}},\ \bibinfo {pages}
  {052319} (\bibinfo {year} {2002})}\BibitemShut {NoStop}%
\bibitem [{\citenamefont {Eckert}\ \emph {et~al.}(2005)\citenamefont {Eckert},
  \citenamefont {Mompart}, \citenamefont {Birkl},\ and\ \citenamefont
  {Lewenstein}}]{eckert2005one}%
  \BibitemOpen
  \bibfield  {author} {\bibinfo {author} {\bibfnamefont {K.}~\bibnamefont
  {Eckert}}, \bibinfo {author} {\bibfnamefont {J.}~\bibnamefont {Mompart}},
  \bibinfo {author} {\bibfnamefont {G.}~\bibnamefont {Birkl}},\ and\ \bibinfo
  {author} {\bibfnamefont {M.}~\bibnamefont {Lewenstein}},\ }\bibfield  {title}
  {\bibinfo {title} {One-and two-dimensional quantum walks in arrays of optical
  traps},\ }\href@noop {} {\bibfield  {journal} {\bibinfo  {journal} {Physical
  Review A}\ }\textbf {\bibinfo {volume} {72}},\ \bibinfo {pages} {012327}
  (\bibinfo {year} {2005})}\BibitemShut {NoStop}%
\bibitem [{\citenamefont {Steffen}\ \emph {et~al.}(2012)\citenamefont
  {Steffen}, \citenamefont {Alberti}, \citenamefont {Alt}, \citenamefont
  {Belmechri}, \citenamefont {Hild}, \citenamefont {Karski}, \citenamefont
  {Widera},\ and\ \citenamefont {Meschede}}]{steffen2012digital}%
  \BibitemOpen
  \bibfield  {author} {\bibinfo {author} {\bibfnamefont {A.}~\bibnamefont
  {Steffen}}, \bibinfo {author} {\bibfnamefont {A.}~\bibnamefont {Alberti}},
  \bibinfo {author} {\bibfnamefont {W.}~\bibnamefont {Alt}}, \bibinfo {author}
  {\bibfnamefont {N.}~\bibnamefont {Belmechri}}, \bibinfo {author}
  {\bibfnamefont {S.}~\bibnamefont {Hild}}, \bibinfo {author} {\bibfnamefont
  {M.}~\bibnamefont {Karski}}, \bibinfo {author} {\bibfnamefont
  {A.}~\bibnamefont {Widera}},\ and\ \bibinfo {author} {\bibfnamefont
  {D.}~\bibnamefont {Meschede}},\ }\bibfield  {title} {\bibinfo {title}
  {Digital atom interferometer with single particle control on a discretized
  space-time geometry},\ }\href@noop {} {\bibfield  {journal} {\bibinfo
  {journal} {Proceedings of the National Academy of Sciences}\ }\textbf
  {\bibinfo {volume} {109}},\ \bibinfo {pages} {9770} (\bibinfo {year}
  {2012})}\BibitemShut {NoStop}%
\bibitem [{\citenamefont {Groh}\ \emph {et~al.}(2016)\citenamefont {Groh},
  \citenamefont {Brakhane}, \citenamefont {Alt}, \citenamefont {Meschede},
  \citenamefont {Asb{\'o}th},\ and\ \citenamefont
  {Alberti}}]{groh2016robustness}%
  \BibitemOpen
  \bibfield  {author} {\bibinfo {author} {\bibfnamefont {T.}~\bibnamefont
  {Groh}}, \bibinfo {author} {\bibfnamefont {S.}~\bibnamefont {Brakhane}},
  \bibinfo {author} {\bibfnamefont {W.}~\bibnamefont {Alt}}, \bibinfo {author}
  {\bibfnamefont {D.}~\bibnamefont {Meschede}}, \bibinfo {author}
  {\bibfnamefont {J.~K.}\ \bibnamefont {Asb{\'o}th}},\ and\ \bibinfo {author}
  {\bibfnamefont {A.}~\bibnamefont {Alberti}},\ }\bibfield  {title} {\bibinfo
  {title} {Robustness of topologically protected edge states in quantum walk
  experiments with neutral atoms},\ }\href@noop {} {\bibfield  {journal}
  {\bibinfo  {journal} {Physical Review A}\ }\textbf {\bibinfo {volume} {94}},\
  \bibinfo {pages} {013620} (\bibinfo {year} {2016})}\BibitemShut {NoStop}%
\bibitem [{\citenamefont {Karski}\ \emph {et~al.}(2009)\citenamefont {Karski},
  \citenamefont {F{\"o}rster}, \citenamefont {Choi}, \citenamefont {Steffen},
  \citenamefont {Alt}, \citenamefont {Meschede},\ and\ \citenamefont
  {Widera}}]{karski2009quantum}%
  \BibitemOpen
  \bibfield  {author} {\bibinfo {author} {\bibfnamefont {M.}~\bibnamefont
  {Karski}}, \bibinfo {author} {\bibfnamefont {L.}~\bibnamefont {F{\"o}rster}},
  \bibinfo {author} {\bibfnamefont {J.-M.}\ \bibnamefont {Choi}}, \bibinfo
  {author} {\bibfnamefont {A.}~\bibnamefont {Steffen}}, \bibinfo {author}
  {\bibfnamefont {W.}~\bibnamefont {Alt}}, \bibinfo {author} {\bibfnamefont
  {D.}~\bibnamefont {Meschede}},\ and\ \bibinfo {author} {\bibfnamefont
  {A.}~\bibnamefont {Widera}},\ }\bibfield  {title} {\bibinfo {title} {Quantum
  walk in position space with single optically trapped atoms},\ }\href@noop {}
  {\bibfield  {journal} {\bibinfo  {journal} {Science}\ }\textbf {\bibinfo
  {volume} {325}},\ \bibinfo {pages} {174} (\bibinfo {year}
  {2009})}\BibitemShut {NoStop}%
\bibitem [{\citenamefont
  {Chandrashekar}(2006)}]{chandrashekar2006implementing}%
  \BibitemOpen
  \bibfield  {author} {\bibinfo {author} {\bibfnamefont {C.}~\bibnamefont
  {Chandrashekar}},\ }\bibfield  {title} {\bibinfo {title} {{Implementing the
  one-dimensional quantum (Hadamard) walk using a Bose-Einstein condensate}},\
  }\href@noop {} {\bibfield  {journal} {\bibinfo  {journal} {Physical Review
  A}\ }\textbf {\bibinfo {volume} {74}},\ \bibinfo {pages} {032307} (\bibinfo
  {year} {2006})}\BibitemShut {NoStop}%
\bibitem [{\citenamefont {Travaglione}\ and\ \citenamefont
  {Milburn}(2002)}]{travaglione2002implementing}%
  \BibitemOpen
  \bibfield  {author} {\bibinfo {author} {\bibfnamefont {B.~C.}\ \bibnamefont
  {Travaglione}}\ and\ \bibinfo {author} {\bibfnamefont {G.~J.}\ \bibnamefont
  {Milburn}},\ }\bibfield  {title} {\bibinfo {title} {Implementing the quantum
  random walk},\ }\href@noop {} {\bibfield  {journal} {\bibinfo  {journal}
  {Physical Review A}\ }\textbf {\bibinfo {volume} {65}},\ \bibinfo {pages}
  {032310} (\bibinfo {year} {2002})}\BibitemShut {NoStop}%
\bibitem [{\citenamefont {Z{\"a}hringer}\ \emph {et~al.}(2010)\citenamefont
  {Z{\"a}hringer}, \citenamefont {Kirchmair}, \citenamefont {Gerritsma},
  \citenamefont {Solano}, \citenamefont {Blatt},\ and\ \citenamefont
  {Roos}}]{zahringer2010realization}%
  \BibitemOpen
  \bibfield  {author} {\bibinfo {author} {\bibfnamefont {F.}~\bibnamefont
  {Z{\"a}hringer}}, \bibinfo {author} {\bibfnamefont {G.}~\bibnamefont
  {Kirchmair}}, \bibinfo {author} {\bibfnamefont {R.}~\bibnamefont
  {Gerritsma}}, \bibinfo {author} {\bibfnamefont {E.}~\bibnamefont {Solano}},
  \bibinfo {author} {\bibfnamefont {R.}~\bibnamefont {Blatt}},\ and\ \bibinfo
  {author} {\bibfnamefont {C.}~\bibnamefont {Roos}},\ }\bibfield  {title}
  {\bibinfo {title} {Realization of a quantum walk with one and two trapped
  ions},\ }\href@noop {} {\bibfield  {journal} {\bibinfo  {journal} {Physical
  Review Letters}\ }\textbf {\bibinfo {volume} {104}},\ \bibinfo {pages}
  {100503} (\bibinfo {year} {2010})}\BibitemShut {NoStop}%
\bibitem [{\citenamefont {Schmitz}\ \emph {et~al.}(2009)\citenamefont
  {Schmitz}, \citenamefont {Matjeschk}, \citenamefont {Schneider},
  \citenamefont {Glueckert}, \citenamefont {Enderlein}, \citenamefont {Huber},\
  and\ \citenamefont {Schaetz}}]{schmitz2009quantum}%
  \BibitemOpen
  \bibfield  {author} {\bibinfo {author} {\bibfnamefont {H.}~\bibnamefont
  {Schmitz}}, \bibinfo {author} {\bibfnamefont {R.}~\bibnamefont {Matjeschk}},
  \bibinfo {author} {\bibfnamefont {C.}~\bibnamefont {Schneider}}, \bibinfo
  {author} {\bibfnamefont {J.}~\bibnamefont {Glueckert}}, \bibinfo {author}
  {\bibfnamefont {M.}~\bibnamefont {Enderlein}}, \bibinfo {author}
  {\bibfnamefont {T.}~\bibnamefont {Huber}},\ and\ \bibinfo {author}
  {\bibfnamefont {T.}~\bibnamefont {Schaetz}},\ }\bibfield  {title} {\bibinfo
  {title} {Quantum walk of a trapped ion in phase space},\ }\href@noop {}
  {\bibfield  {journal} {\bibinfo  {journal} {Physical Review Letters}\
  }\textbf {\bibinfo {volume} {103}},\ \bibinfo {pages} {090504} (\bibinfo
  {year} {2009})}\BibitemShut {NoStop}%
\bibitem [{\citenamefont {Perets}\ \emph {et~al.}(2008)\citenamefont {Perets},
  \citenamefont {Lahini}, \citenamefont {Pozzi}, \citenamefont {Sorel},
  \citenamefont {Morandotti},\ and\ \citenamefont
  {Silberberg}}]{perets2008realization}%
  \BibitemOpen
  \bibfield  {author} {\bibinfo {author} {\bibfnamefont {H.~B.}\ \bibnamefont
  {Perets}}, \bibinfo {author} {\bibfnamefont {Y.}~\bibnamefont {Lahini}},
  \bibinfo {author} {\bibfnamefont {F.}~\bibnamefont {Pozzi}}, \bibinfo
  {author} {\bibfnamefont {M.}~\bibnamefont {Sorel}}, \bibinfo {author}
  {\bibfnamefont {R.}~\bibnamefont {Morandotti}},\ and\ \bibinfo {author}
  {\bibfnamefont {Y.}~\bibnamefont {Silberberg}},\ }\bibfield  {title}
  {\bibinfo {title} {Realization of quantum walks with negligible decoherence
  in waveguide lattices},\ }\href@noop {} {\bibfield  {journal} {\bibinfo
  {journal} {Physical Review Letters}\ }\textbf {\bibinfo {volume} {100}},\
  \bibinfo {pages} {170506} (\bibinfo {year} {2008})}\BibitemShut {NoStop}%
\bibitem [{\citenamefont {Peruzzo}\ \emph {et~al.}(2010)\citenamefont
  {Peruzzo}, \citenamefont {Lobino}, \citenamefont {Matthews}, \citenamefont
  {Matsuda}, \citenamefont {Politi}, \citenamefont {Poulios}, \citenamefont
  {Zhou}, \citenamefont {Lahini}, \citenamefont {Ismail}, \citenamefont
  {W{\"o}rhoff} \emph {et~al.}}]{peruzzo2010quantum}%
  \BibitemOpen
  \bibfield  {author} {\bibinfo {author} {\bibfnamefont {A.}~\bibnamefont
  {Peruzzo}}, \bibinfo {author} {\bibfnamefont {M.}~\bibnamefont {Lobino}},
  \bibinfo {author} {\bibfnamefont {J.~C.}\ \bibnamefont {Matthews}}, \bibinfo
  {author} {\bibfnamefont {N.}~\bibnamefont {Matsuda}}, \bibinfo {author}
  {\bibfnamefont {A.}~\bibnamefont {Politi}}, \bibinfo {author} {\bibfnamefont
  {K.}~\bibnamefont {Poulios}}, \bibinfo {author} {\bibfnamefont {X.-Q.}\
  \bibnamefont {Zhou}}, \bibinfo {author} {\bibfnamefont {Y.}~\bibnamefont
  {Lahini}}, \bibinfo {author} {\bibfnamefont {N.}~\bibnamefont {Ismail}},
  \bibinfo {author} {\bibfnamefont {K.}~\bibnamefont {W{\"o}rhoff}}, \emph
  {et~al.},\ }\bibfield  {title} {\bibinfo {title} {Quantum walks of correlated
  photons},\ }\href@noop {} {\bibfield  {journal} {\bibinfo  {journal}
  {Science}\ }\textbf {\bibinfo {volume} {329}},\ \bibinfo {pages} {1500}
  (\bibinfo {year} {2010})}\BibitemShut {NoStop}%
\bibitem [{\citenamefont {Cardano}\ \emph {et~al.}(2017)\citenamefont
  {Cardano}, \citenamefont {D?Errico}, \citenamefont {Dauphin}, \citenamefont
  {Maffei}, \citenamefont {Piccirillo}, \citenamefont {de~Lisio}, \citenamefont
  {De~Filippis}, \citenamefont {Cataudella}, \citenamefont {Santamato},
  \citenamefont {Marrucci} \emph {et~al.}}]{cardano2017detection}%
  \BibitemOpen
  \bibfield  {author} {\bibinfo {author} {\bibfnamefont {F.}~\bibnamefont
  {Cardano}}, \bibinfo {author} {\bibfnamefont {A.}~\bibnamefont {D?Errico}},
  \bibinfo {author} {\bibfnamefont {A.}~\bibnamefont {Dauphin}}, \bibinfo
  {author} {\bibfnamefont {M.}~\bibnamefont {Maffei}}, \bibinfo {author}
  {\bibfnamefont {B.}~\bibnamefont {Piccirillo}}, \bibinfo {author}
  {\bibfnamefont {C.}~\bibnamefont {de~Lisio}}, \bibinfo {author}
  {\bibfnamefont {G.}~\bibnamefont {De~Filippis}}, \bibinfo {author}
  {\bibfnamefont {V.}~\bibnamefont {Cataudella}}, \bibinfo {author}
  {\bibfnamefont {E.}~\bibnamefont {Santamato}}, \bibinfo {author}
  {\bibfnamefont {L.}~\bibnamefont {Marrucci}}, \emph {et~al.},\ }\bibfield
  {title} {\bibinfo {title} {Detection of {Zak} phases and topological
  invariants in a chiral quantum walk of twisted photons},\ }\href@noop {}
  {\bibfield  {journal} {\bibinfo  {journal} {Nature Communications}\ }\textbf
  {\bibinfo {volume} {8}},\ \bibinfo {pages} {15516} (\bibinfo {year}
  {2017})}\BibitemShut {NoStop}%
\bibitem [{\citenamefont {Chen}\ \emph {et~al.}(2018)\citenamefont {Chen},
  \citenamefont {Ding}, \citenamefont {Qin}, \citenamefont {He}, \citenamefont
  {Luo}, \citenamefont {Chen}, \citenamefont {Liu}, \citenamefont {Wang},
  \citenamefont {Zhang}, \citenamefont {Li} \emph
  {et~al.}}]{chen2018observation}%
  \BibitemOpen
  \bibfield  {author} {\bibinfo {author} {\bibfnamefont {C.}~\bibnamefont
  {Chen}}, \bibinfo {author} {\bibfnamefont {X.}~\bibnamefont {Ding}}, \bibinfo
  {author} {\bibfnamefont {J.}~\bibnamefont {Qin}}, \bibinfo {author}
  {\bibfnamefont {Y.}~\bibnamefont {He}}, \bibinfo {author} {\bibfnamefont
  {Y.-H.}\ \bibnamefont {Luo}}, \bibinfo {author} {\bibfnamefont {M.-C.}\
  \bibnamefont {Chen}}, \bibinfo {author} {\bibfnamefont {C.}~\bibnamefont
  {Liu}}, \bibinfo {author} {\bibfnamefont {X.-L.}\ \bibnamefont {Wang}},
  \bibinfo {author} {\bibfnamefont {W.-J.}\ \bibnamefont {Zhang}}, \bibinfo
  {author} {\bibfnamefont {H.}~\bibnamefont {Li}}, \emph {et~al.},\ }\bibfield
  {title} {\bibinfo {title} {Observation of topologically protected edge states
  in a photonic two-dimensional quantum walk},\ }\href@noop {} {\bibfield
  {journal} {\bibinfo  {journal} {Physical Review Letters}\ }\textbf {\bibinfo
  {volume} {121}},\ \bibinfo {pages} {100502} (\bibinfo {year}
  {2018})}\BibitemShut {NoStop}%
\bibitem [{\citenamefont {Tang}\ \emph {et~al.}(2018)\citenamefont {Tang},
  \citenamefont {Lin}, \citenamefont {Feng}, \citenamefont {Chen},
  \citenamefont {Gao}, \citenamefont {Sun}, \citenamefont {Wang}, \citenamefont
  {Lai}, \citenamefont {Xu}, \citenamefont {Wang} \emph
  {et~al.}}]{tang2018experimental}%
  \BibitemOpen
  \bibfield  {author} {\bibinfo {author} {\bibfnamefont {H.}~\bibnamefont
  {Tang}}, \bibinfo {author} {\bibfnamefont {X.-F.}\ \bibnamefont {Lin}},
  \bibinfo {author} {\bibfnamefont {Z.}~\bibnamefont {Feng}}, \bibinfo {author}
  {\bibfnamefont {J.-Y.}\ \bibnamefont {Chen}}, \bibinfo {author}
  {\bibfnamefont {J.}~\bibnamefont {Gao}}, \bibinfo {author} {\bibfnamefont
  {K.}~\bibnamefont {Sun}}, \bibinfo {author} {\bibfnamefont {C.-Y.}\
  \bibnamefont {Wang}}, \bibinfo {author} {\bibfnamefont {P.-C.}\ \bibnamefont
  {Lai}}, \bibinfo {author} {\bibfnamefont {X.-Y.}\ \bibnamefont {Xu}},
  \bibinfo {author} {\bibfnamefont {Y.}~\bibnamefont {Wang}}, \emph {et~al.},\
  }\bibfield  {title} {\bibinfo {title} {Experimental two-dimensional quantum
  walk on a photonic chip},\ }\href@noop {} {\bibfield  {journal} {\bibinfo
  {journal} {Sci. Adv.}\ }\textbf {\bibinfo {volume} {4}},\ \bibinfo {pages}
  {eaat3174} (\bibinfo {year} {2018})}\BibitemShut {NoStop}%
\bibitem [{\citenamefont {Poulios}\ \emph {et~al.}(2014)\citenamefont
  {Poulios}, \citenamefont {Keil}, \citenamefont {Fry}, \citenamefont
  {Meinecke}, \citenamefont {Matthews}, \citenamefont {Politi}, \citenamefont
  {Lobino}, \citenamefont {Gr{\"a}fe}, \citenamefont {Heinrich}, \citenamefont
  {Nolte} \emph {et~al.}}]{poulios2014quantum}%
  \BibitemOpen
  \bibfield  {author} {\bibinfo {author} {\bibfnamefont {K.}~\bibnamefont
  {Poulios}}, \bibinfo {author} {\bibfnamefont {R.}~\bibnamefont {Keil}},
  \bibinfo {author} {\bibfnamefont {D.}~\bibnamefont {Fry}}, \bibinfo {author}
  {\bibfnamefont {J.~D.}\ \bibnamefont {Meinecke}}, \bibinfo {author}
  {\bibfnamefont {J.~C.}\ \bibnamefont {Matthews}}, \bibinfo {author}
  {\bibfnamefont {A.}~\bibnamefont {Politi}}, \bibinfo {author} {\bibfnamefont
  {M.}~\bibnamefont {Lobino}}, \bibinfo {author} {\bibfnamefont
  {M.}~\bibnamefont {Gr{\"a}fe}}, \bibinfo {author} {\bibfnamefont
  {M.}~\bibnamefont {Heinrich}}, \bibinfo {author} {\bibfnamefont
  {S.}~\bibnamefont {Nolte}}, \emph {et~al.},\ }\bibfield  {title} {\bibinfo
  {title} {Quantum walks of correlated photon pairs in two-dimensional
  waveguide arrays},\ }\href@noop {} {\bibfield  {journal} {\bibinfo  {journal}
  {Physical Review Letters}\ }\textbf {\bibinfo {volume} {112}},\ \bibinfo
  {pages} {143604} (\bibinfo {year} {2014})}\BibitemShut {NoStop}%
\bibitem [{\citenamefont {Schreiber}\ \emph {et~al.}(2010)\citenamefont
  {Schreiber}, \citenamefont {Cassemiro}, \citenamefont {Poto{\v{c}}ek},
  \citenamefont {G{\'a}bris}, \citenamefont {Mosley}, \citenamefont
  {Andersson}, \citenamefont {Jex},\ and\ \citenamefont
  {Silberhorn}}]{schreiber2010photons}%
  \BibitemOpen
  \bibfield  {author} {\bibinfo {author} {\bibfnamefont {A.}~\bibnamefont
  {Schreiber}}, \bibinfo {author} {\bibfnamefont {K.~N.}\ \bibnamefont
  {Cassemiro}}, \bibinfo {author} {\bibfnamefont {V.}~\bibnamefont
  {Poto{\v{c}}ek}}, \bibinfo {author} {\bibfnamefont {A.}~\bibnamefont
  {G{\'a}bris}}, \bibinfo {author} {\bibfnamefont {P.~J.}\ \bibnamefont
  {Mosley}}, \bibinfo {author} {\bibfnamefont {E.}~\bibnamefont {Andersson}},
  \bibinfo {author} {\bibfnamefont {I.}~\bibnamefont {Jex}},\ and\ \bibinfo
  {author} {\bibfnamefont {C.}~\bibnamefont {Silberhorn}},\ }\bibfield  {title}
  {\bibinfo {title} {Photons walking the line: a quantum walk with adjustable
  coin operations},\ }\href@noop {} {\bibfield  {journal} {\bibinfo  {journal}
  {Physical Review Letters}\ }\textbf {\bibinfo {volume} {104}},\ \bibinfo
  {pages} {050502} (\bibinfo {year} {2010})}\BibitemShut {NoStop}%
\bibitem [{\citenamefont {Raizen}(1999)}]{Raizen1999}%
  \BibitemOpen
  \bibfield  {author} {\bibinfo {author} {\bibfnamefont {M.~G.}\ \bibnamefont
  {Raizen}},\ }\bibfield  {title} {\bibinfo {title} {Quantum chaos with cold
  atoms},\ }\href@noop {} {\bibfield  {journal} {\bibinfo  {journal} {Advances
  in Atomic, Molecular, and Optical Physics}\ }\textbf {\bibinfo {volume}
  {41}},\ \bibinfo {pages} {43} (\bibinfo {year} {1999})}\BibitemShut
  {NoStop}%
\bibitem [{\citenamefont {Sadgrove}\ and\ \citenamefont
  {Wimberger}(2011)}]{SW2011}%
  \BibitemOpen
  \bibfield  {author} {\bibinfo {author} {\bibfnamefont {M.}~\bibnamefont
  {Sadgrove}}\ and\ \bibinfo {author} {\bibfnamefont {S.}~\bibnamefont
  {Wimberger}},\ }\bibfield  {title} {\bibinfo {title} {{A pseudo-classical
  method for the atom-optics kicked rotor: from theory to experiment and
  back}},\ }\href@noop {} {\bibfield  {journal} {\bibinfo  {journal} {Adv. At.
  Mol. Opt. Phys.}\ }\textbf {\bibinfo {volume} {60}},\ \bibinfo {pages} {315}
  (\bibinfo {year} {2011})}\BibitemShut {NoStop}%
\bibitem [{\citenamefont {Wimberger}\ \emph {et~al.}(2003)\citenamefont
  {Wimberger}, \citenamefont {Guarneri},\ and\ \citenamefont
  {Fishman}}]{WGF2003}%
  \BibitemOpen
  \bibfield  {author} {\bibinfo {author} {\bibfnamefont {S.}~\bibnamefont
  {Wimberger}}, \bibinfo {author} {\bibfnamefont {I.}~\bibnamefont
  {Guarneri}},\ and\ \bibinfo {author} {\bibfnamefont {S.}~\bibnamefont
  {Fishman}},\ }\bibfield  {title} {\bibinfo {title} {{Quantum resonances and
  decoherence for delta-kicked atoms}},\ }\href
  {http://stacks.iop.org/0951-7715/16/i=4/a=312} {\bibfield  {journal}
  {\bibinfo  {journal} {Nonlinearity}\ }\textbf {\bibinfo {volume} {16}},\
  \bibinfo {pages} {1381} (\bibinfo {year} {2003})}\BibitemShut {NoStop}%
\bibitem [{\citenamefont {Ni}\ \emph {et~al.}(2016)\citenamefont {Ni},
  \citenamefont {Lam}, \citenamefont {Dadras}, \citenamefont {Borunda},
  \citenamefont {Wimberger},\ and\ \citenamefont {Summy}}]{Ni2016}%
  \BibitemOpen
  \bibfield  {author} {\bibinfo {author} {\bibfnamefont {J.}~\bibnamefont
  {Ni}}, \bibinfo {author} {\bibfnamefont {W.~K.}\ \bibnamefont {Lam}},
  \bibinfo {author} {\bibfnamefont {S.}~\bibnamefont {Dadras}}, \bibinfo
  {author} {\bibfnamefont {M.~F.}\ \bibnamefont {Borunda}}, \bibinfo {author}
  {\bibfnamefont {S.}~\bibnamefont {Wimberger}},\ and\ \bibinfo {author}
  {\bibfnamefont {G.~S.}\ \bibnamefont {Summy}},\ }\bibfield  {title} {\bibinfo
  {title} {Initial-state dependence of a quantum resonance ratchet},\ }\href
  {https://doi.org/10.1103/PhysRevA.94.043620} {\bibfield  {journal} {\bibinfo
  {journal} {Phys. Rev. A}\ }\textbf {\bibinfo {volume} {94}},\ \bibinfo
  {pages} {043620} (\bibinfo {year} {2016})}\BibitemShut {NoStop}%
\bibitem [{\citenamefont {Ni}\ \emph {et~al.}(2017)\citenamefont {Ni},
  \citenamefont {Dadras}, \citenamefont {Lam}, \citenamefont {Shrestha},
  \citenamefont {Sadgrove}, \citenamefont {Wimberger},\ and\ \citenamefont
  {Summy}}]{Ni2017}%
  \BibitemOpen
  \bibfield  {author} {\bibinfo {author} {\bibfnamefont {J.}~\bibnamefont
  {Ni}}, \bibinfo {author} {\bibfnamefont {S.}~\bibnamefont {Dadras}}, \bibinfo
  {author} {\bibfnamefont {W.}~\bibnamefont {Lam}}, \bibinfo {author}
  {\bibfnamefont {R.}~\bibnamefont {Shrestha}}, \bibinfo {author}
  {\bibfnamefont {M.}~\bibnamefont {Sadgrove}}, \bibinfo {author}
  {\bibfnamefont {S.}~\bibnamefont {Wimberger}},\ and\ \bibinfo {author}
  {\bibfnamefont {G.~S.}\ \bibnamefont {Summy}},\ }\bibfield  {title} {\bibinfo
  {title} {Hamiltonian ratchets with ultra-cold atoms},\ }\href
  {https://doi.org/10.1002/andp.201600335} {\bibfield  {journal} {\bibinfo
  {journal} {Annalen der Physik}\ }\textbf {\bibinfo {volume} {529}},\ \bibinfo
  {pages} {1600335} (\bibinfo {year} {2017})}\BibitemShut {NoStop}%
\bibitem [{\citenamefont {Sadgrove}\ \emph {et~al.}(2007)\citenamefont
  {Sadgrove}, \citenamefont {Horikoshi}, \citenamefont {Sekimura},\ and\
  \citenamefont {Nakagawa}}]{Mark2007}%
  \BibitemOpen
  \bibfield  {author} {\bibinfo {author} {\bibfnamefont {M.}~\bibnamefont
  {Sadgrove}}, \bibinfo {author} {\bibfnamefont {M.}~\bibnamefont {Horikoshi}},
  \bibinfo {author} {\bibfnamefont {T.}~\bibnamefont {Sekimura}},\ and\
  \bibinfo {author} {\bibfnamefont {K.}~\bibnamefont {Nakagawa}},\ }\bibfield
  {title} {\bibinfo {title} {{Rectified Momentum Transport for a Kicked
  Bose-Einstein Condensate}},\ }\href
  {https://doi.org/10.1103/PhysRevLett.99.043002} {\bibfield  {journal}
  {\bibinfo  {journal} {Phys. Rev. Lett.}\ }\textbf {\bibinfo {volume} {99}},\
  \bibinfo {pages} {043002} (\bibinfo {year} {2007})}\BibitemShut {NoStop}%
\bibitem [{\citenamefont {Dana}\ \emph {et~al.}(2008)\citenamefont {Dana},
  \citenamefont {Ramareddy}, \citenamefont {Talukdar},\ and\ \citenamefont
  {Summy}}]{Gil2008}%
  \BibitemOpen
  \bibfield  {author} {\bibinfo {author} {\bibfnamefont {I.}~\bibnamefont
  {Dana}}, \bibinfo {author} {\bibfnamefont {V.}~\bibnamefont {Ramareddy}},
  \bibinfo {author} {\bibfnamefont {I.}~\bibnamefont {Talukdar}},\ and\
  \bibinfo {author} {\bibfnamefont {G.~S.}\ \bibnamefont {Summy}},\ }\bibfield
  {title} {\bibinfo {title} {{Experimental Realization of Quantum-Resonance
  Ratchets at Arbitrary Quasimomenta}},\ }\href
  {https://doi.org/10.1103/PhysRevLett.100.024103} {\bibfield  {journal}
  {\bibinfo  {journal} {Phys. Rev. Lett.}\ }\textbf {\bibinfo {volume} {100}},\
  \bibinfo {pages} {024103} (\bibinfo {year} {2008})}\BibitemShut {NoStop}%
\bibitem [{\citenamefont {Summy}\ and\ \citenamefont
  {Wimberger}(2016)}]{SW2016}%
  \BibitemOpen
  \bibfield  {author} {\bibinfo {author} {\bibfnamefont {G.~S.}\ \bibnamefont
  {Summy}}\ and\ \bibinfo {author} {\bibfnamefont {S.}~\bibnamefont
  {Wimberger}},\ }\bibfield  {title} {\bibinfo {title} {{Quantum random walk of
  a Bose-Einstein condensate in momentum space}},\ }\href
  {https://doi.org/10.1103/PhysRevA.93.023638} {\bibfield  {journal} {\bibinfo
  {journal} {Phys. Rev. A}\ }\textbf {\bibinfo {volume} {93}},\ \bibinfo
  {pages} {023638} (\bibinfo {year} {2016})}\BibitemShut {NoStop}%
\bibitem [{\citenamefont {Groiseau}(2017)}]{Groiseau2017}%
  \BibitemOpen
  \bibfield  {author} {\bibinfo {author} {\bibfnamefont {C.}~\bibnamefont
  {Groiseau}},\ }\emph {\bibinfo {title} {Discrete-time quantum walks in
  momentum space}},\ \href@noop {} {Master's thesis},\ \bibinfo  {school}
  {Heidelberg University} (\bibinfo {year} {2017})\BibitemShut {NoStop}%
\bibitem [{\citenamefont {Groiseau}\ \emph {et~al.}(2018)\citenamefont
  {Groiseau}, \citenamefont {Gresch},\ and\ \citenamefont
  {Wimberger}}]{Groiseau2018}%
  \BibitemOpen
  \bibfield  {author} {\bibinfo {author} {\bibfnamefont {C.}~\bibnamefont
  {Groiseau}}, \bibinfo {author} {\bibfnamefont {A.}~\bibnamefont {Gresch}},\
  and\ \bibinfo {author} {\bibfnamefont {S.}~\bibnamefont {Wimberger}},\
  }\bibfield  {title} {\bibinfo {title} {Quantum walks of kicked
  {Bose-Einstein} condensates},\ }\href
  {https://doi.org/10.1088/1751-8121/aac5d5} {\bibfield  {journal} {\bibinfo
  {journal} {Journal of Physics A: Mathematical and Theoretical}\ }\textbf
  {\bibinfo {volume} {51}},\ \bibinfo {pages} {275301} (\bibinfo {year}
  {2018})}\BibitemShut {NoStop}%
\bibitem [{\citenamefont {Delone}\ and\ \citenamefont
  {Krainov}(1999)}]{Delone1999}%
  \BibitemOpen
  \bibfield  {author} {\bibinfo {author} {\bibfnamefont {N.~B.}\ \bibnamefont
  {Delone}}\ and\ \bibinfo {author} {\bibfnamefont {V.~P.}\ \bibnamefont
  {Krainov}},\ }\bibfield  {title} {\bibinfo {title} {{AC Stark} shift of
  atomic energy levels},\ }\href
  {https://doi.org/10.1070/pu1999v042n07abeh000557} {\bibfield  {journal}
  {\bibinfo  {journal} {Physics-Uspekhi}\ }\textbf {\bibinfo {volume} {42}},\
  \bibinfo {pages} {669} (\bibinfo {year} {1999})}\BibitemShut {NoStop}%
\bibitem [{\citenamefont {Delvecchio}\ \emph {et~al.}(2020)\citenamefont
  {Delvecchio}, \citenamefont {Groiseau}, \citenamefont {Petiziol},
  \citenamefont {Summy},\ and\ \citenamefont {Wimberger}}]{Delvecchio2020}%
  \BibitemOpen
  \bibfield  {author} {\bibinfo {author} {\bibfnamefont {M.}~\bibnamefont
  {Delvecchio}}, \bibinfo {author} {\bibfnamefont {C.}~\bibnamefont
  {Groiseau}}, \bibinfo {author} {\bibfnamefont {F.}~\bibnamefont {Petiziol}},
  \bibinfo {author} {\bibfnamefont {G.~S.}\ \bibnamefont {Summy}},\ and\
  \bibinfo {author} {\bibfnamefont {S.}~\bibnamefont {Wimberger}},\ }\bibfield
  {title} {\bibinfo {title} {Quantum search with a continuous-time quantum walk
  in momentum space},\ }\href {https://doi.org/10.1088/1361-6455/ab63ad}
  {\bibfield  {journal} {\bibinfo  {journal} {Journal of Physics B: Atomic,
  Molecular and Optical Physics}\ }\textbf {\bibinfo {volume} {53}},\ \bibinfo
  {pages} {065301} (\bibinfo {year} {2020})}\BibitemShut {NoStop}%
\bibitem [{\citenamefont {Groiseau}\ \emph {et~al.}(2019)\citenamefont
  {Groiseau}, \citenamefont {Wagner}, \citenamefont {Summy},\ and\
  \citenamefont {Wimberger}}]{Groiseau2019}%
  \BibitemOpen
  \bibfield  {author} {\bibinfo {author} {\bibfnamefont {C.}~\bibnamefont
  {Groiseau}}, \bibinfo {author} {\bibfnamefont {A.}~\bibnamefont {Wagner}},
  \bibinfo {author} {\bibfnamefont {G.~S.}\ \bibnamefont {Summy}},\ and\
  \bibinfo {author} {\bibfnamefont {S.}~\bibnamefont {Wimberger}},\ }\bibfield
  {title} {\bibinfo {title} {Impact of lattice vibrations on the dynamics of a
  spinor atom-optics kicked rotor},\ }\href
  {https://doi.org/10.3390/condmat4010010} {\bibfield  {journal} {\bibinfo
  {journal} {Condensed Matter}\ }\textbf {\bibinfo {volume} {4}},\ \bibinfo
  {pages} {10} (\bibinfo {year} {2019})}\BibitemShut {NoStop}%
\bibitem [{\citenamefont {Abramowitz}\ and\ \citenamefont
  {Stegun}(1964)}]{abramowitz1964handbook}%
  \BibitemOpen
  \bibfield  {author} {\bibinfo {author} {\bibfnamefont {M.}~\bibnamefont
  {Abramowitz}}\ and\ \bibinfo {author} {\bibfnamefont {I.~A.}\ \bibnamefont
  {Stegun}},\ }\href@noop {} {\emph {\bibinfo {title} {Handbook of mathematical
  functions with formulas, graphs, and mathematical tables}}},\ Vol.~\bibinfo
  {volume} {55}\ (\bibinfo  {publisher} {US Government printing office},\
  \bibinfo {year} {1964})\BibitemShut {NoStop}%
\end{thebibliography}

\end{document}